\documentclass[a4paper,fleqn]{cas-sc}
\usepackage[numbers]{natbib}
\usepackage{float}
\usepackage{graphicx}
\usepackage{amsmath}
\usepackage{amssymb}
\usepackage{enumitem}
\usepackage{amsfonts}
\usepackage[many]{tcolorbox}
\usepackage{diagbox}
\usepackage{balance}
\usepackage{geometry}
\usepackage{setspace}
\usepackage{color}
\usepackage{blindtext}
\usepackage[export]{adjustbox}
\usepackage{multicol}
\usepackage[utf8]{inputenc}
\everymath{\displaystyle}
\usepackage[figuresright]{rotating}
\usepackage{array}
\usepackage{kantlipsum}
\usepackage{gensymb}
\usepackage{xpatch}
\usepackage{enumerate}
\usepackage{lipsum}
\usepackage{multirow} 
\usepackage{lineno}
\usepackage{nccmath}
\usepackage{amsbsy}
\usepackage{ragged2e}
\usepackage{subcaption}
\usepackage{etoolbox}
\usepackage{siunitx}
\usepackage{mathtools}
\usepackage{fullpage,anysize,esint,relsize, commath}
\usepackage{pifont}
\usepackage{listings}
\usepackage{cprotect}
\usepackage{booktabs}
\usepackage{longtable}
\usepackage{soul}
\usepackage{isomath}
\usepackage{graphics}
\usepackage{algorithm}
\usepackage{algorithmic}
\usepackage[dvipsnames]{xcolor}
\usepackage[T1]{fontenc}
\DeclareUnicodeCharacter{FF08}{(}
\DeclareUnicodeCharacter{FF09}{)}
\DeclareMathAlphabet{\mathpzc}{OT1}{pzc}{m}{it}
\DeclareMathAlphabet{\mathbbm}{U}{bbm}{m}{n}
\def\tsc#1{\csdef{#1}{\textsc{\lowercase{#1}}\xspace}}
\tsc{WGM}
\tsc{QE}
\tsc{EP}
\tsc{PMS}
\tsc{BEC}
\tsc{DE}

\begin{document}
\let\WriteBookmarks\relax
\def\floatpagepagefraction{1}
\def\textpagefraction{.001}

\renewcommand\floatpagefraction{.9}
\renewcommand\topfraction{.9}
\renewcommand\bottomfraction{.9}
\renewcommand\textfraction{.1}   
\setcounter{totalnumber}{50}
\setcounter{topnumber}{50}
\setcounter{bottomnumber}{50}

\title [mode = title]{A Novel Active Learning Approach to Label One Million Unknown Malware Variants}  
\author[1]{Ahmed Bensaoud}
\cormark[1]
\ead{abensaou@uccs.edu}
\author[1]{Jugal Kalita}
\ead{jkalita@uccs.edu}
\address[1]{Department of Computer Science, University of Colorado Colorado Springs, USA}

\begin{abstract}
Active learning for classification seeks to reduce the cost of labeling samples by finding unlabeled examples about which the current model is least certain and sending them to an annotator/expert to label. Bayesian theory can provide a probabilistic view of deep neural network models by asserting a prior distribution over model parameters and estimating the uncertainties by posterior distribution over these parameters. This paper proposes two novel active learning approaches to label one million malware examples belonging to different unknown modern malware families. The first model is Inception-V4+PCA combined with several support vector machine (SVM) algorithms (UTSVM, PSVM, SVM-GSU, TBSVM). The second model is Vision Transformer based Bayesian Neural Networks ViT-BNN. Our proposed ViT-BNN is a state-of-the-art  active learning approach that differs from current methods and can apply to any particular task. The experiments demonstrate that the ViT-BNN is more stable and robust in handling uncertainty.
\end{abstract}


\maketitle
\section{Introduction}

The goal of malware purveyors is to create nuisance, instill fear and extort users. Cybersecurity Ventures expects the global cost of cyber-crime to increase by 15 percent per year, rising to \$10.5 trillion in 2025\footnote{\url{https://cybersecurityventures.com/cybercrime-damages-6-trillion-by-2021/}}. From January to April 2023, a total of 22 new vulnerabilities that are associated with malware have been identified \footnote{\url{https://cybersecurityworks.com/ransomware}}. The fact is, malware poses persistent and substantial threat to people and organizations. Malware has been lucrative for hackers, who are constantly looking for new and different methods for extortion. Malware is heading toward becoming a much bigger problem with continued proliferation of connected devices that are likely to become victims to attacks. What motivates malware hackers to become even more creative in their attacks, making demands and asking for millions of dollars in payments is that many companies agree to pay the ransom without disclosing the attack.

To create and train better approaches to protect systems and devices, a new malware dataset is needed to train detection mechanisms to discriminate malware from other types of malware as well as benign files \cite{bensaoud2024survey,bensaoud2024cnn,bensaoud2022deep}. Active learning (AL) works with a small amount of labeled data examples and a large amount of unlabeled data. Active learning is an approach to choose which unlabeled data points we should ask a human to label in order to get the best learning gains in the future. In this context, AL identifies \textbf{high-value} samples data points whose accurate labeling would significantly enhance classification performance, thereby improving the overall efficiency and effectiveness of the learning process.

\textbf{High-value} samples play a crucial role in the success of active learning methodologies. These samples are identified as the most informative instances within the dataset, often representing edge cases, rare patterns, or highly complex examples that challenge the model's predictive capability. By focusing annotation efforts on high-value samples, active learning ensures that each labeled data point contributes disproportionately to the improvement of the model. This targeted approach allows for faster convergence of the learning process while reducing the overall annotation workload, making it an indispensable strategy in scenarios with large-scale unlabeled datasets, such as malware classification.

Active learning first takes a small amount of labeled data and provides that to a machine learning model (maybe deep learning)  which tries to label the rest of the data automatically. The difference between traditional passive learning and active learning is clearly shown in Figures 1 and 2. The initial classifier is trained on the small set of labeled data, and improved repeatedly, with a limited amount of human involvement as described below. The by-product is a large labeled dataset.  

The classifier identifies a subset of the unlabeled data that it is most uncertain about and presents these \textbf{high-value} samples for human annotation. Using these uncertain samples, the classifier can efficiently improve its performance through iterative human feedback. This targeted approach to sample selection not only enhances the learning process but also reduces the amount of human effort required for annotation. After several rounds of annotation and retraining, the system becomes less confused and is capable of labeling the remaining data autonomously. Consequently, active learning can reduce the total amount of human effort required for annotation by 98.9\% to 99.9\%.\\
In summary, we make the following contributions:
\begin{itemize}
\item We present a novel Vision Transformer based neural network with Bayesian Neural Networks to handle the uncertainty in classification of unknown malware examples.

\item We also evaluate four different support vector machine (SVM) algorithms with a CNN network called Inception-V4, along with the feature reduction approach called Principal Component Analysis (PCA) to handle the uncertainty. The SVM algorithms are UTSVM \cite{liang2022uncertainty}, PSVM \cite{zhang2012power}, SVM-GSU \cite{tzelepis2017linear}, and TBSVM \cite{shao2011improvements}.
\end{itemize}

The remainder of this paper is organized as follows: Section 2 presents related work. Section 3 reviews acquisition functions. Section 4 describes the proposed methodology including the dataset to be labeled. Finally, experimental results for two models are presented in Section 5.

\section{Related work}
\citet{lin2021active} proposed an active learning approach using a small amount of labeled samples, and malicious mislabeled samples, including generated adversarial examples. The model gave weak results in term of accuracy, F1-score, and Area Under the Receiver Operating Characteristic Curve (AUC-ROC). \citet{Boukela1} proposed a new active learning approach using unlabeled examples and a KNN-based query function to help the labeling process, to detect unknown attacks. \citet{teye2018bayesian} designed Monte Carlo (MC) batch normalization (MCBN) to estimate the uncertainty in networks with batch normalization. In addition, they claimed that batch normalization can be envisioned as an approximate Bayesian model. \citet{foong2020expressiveness} reviewed the mean-field Gaussian variational inference and MC dropout, and discovered that both approaches could be used to represent uncertainty in BNNs. \citet{zhang2019cyclical} proposed a cyclical stochastic gradient (SG) and Markov chain Monte Carlo (MCMC) approach to calculate the posterior over the weights of deep neural networks. Unfortunately, despite their model reducing the computational complexity using a mini-batch of the dataset at each iteration to update the model parameters, the model increased the uncertainty in classification.

\citet{louizos2017multiplicative} implemented a stochastic gradient variational inference approach to calculate the posterior distributions over the weights of DNNs. \citet{ebrahimi2019uncertainty} built an uncertainty-guided continual approach to show uncertainty for each weight to control the adjustment in the parameters of a Bayesian neural network. \citet{choi2019gaussian} proposed a Gaussian-YOLOv3 approach to predict the localization uncertainty through the detection process. Their method significantly reduced the False Positive (FP) rate and increased the True Positive (TP) rate. \citet{li2024deep} proposed an algorithm, NoiseStability, for uncertainty estimation in deep active learning by measuring output deviations when model parameters are perturbed by noise. \citet{hu2024uncertainty} proposed active developmental learning (ADL), a task enabling models to progressively enhance their ability to detect unfamiliar and unknown classes by actively selecting what to learn. Their uncertainty-driven method (UADL) effectively identifies and handles unfamiliar known classes and unknown classes through a two-stage process. \citet{HUANG2023102282} proposed a weighting filter (W-filter) tailored for object detection, and resample the data with uncertain labels from the frequency domain to alleviate the problem of data bias. \citet{chen2023semi} proposed a stability aggregation score to aggregate the classification and localization stability scores and dynamic adaptive threshold to adjust the entries for each category label to alleviate the class imbalance problem.  The goal was to measure the uncertainty of an unlabeled image from classification stability and localization stability. \citet{li2024active} explored the integration of active learning (AL) into data quality control (DQC) processes, emphasizing its potential to reduce the data labeling burden on domain experts while addressing common data quality issues such as anomalies, scarcity, and imbalance in machine learning (ML) contexts. They reviewed existing data quality challenges and AL techniques, presenting two scenarios for AL application in anomaly detection tasks within DQC systems and highlighting AL’s ability to enhance data quality assessment through human-in-the-loop approaches. 

\begin{figure*}[!ht]
	\centering
		\includegraphics[width=16cm, height=9cm]{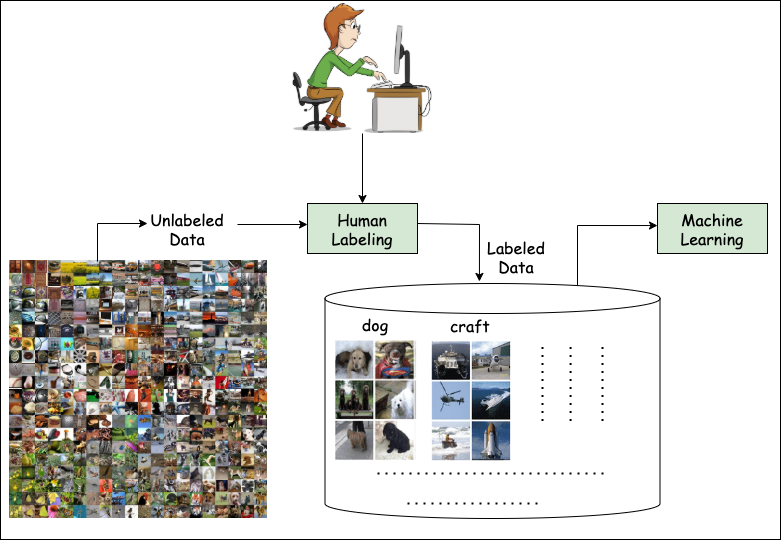}
	  \caption{Passive Learning}\label{fig1}
\end{figure*}

\begin{figure*}[!ht]
	\centering
		\includegraphics[width=16cm, height=9cm]{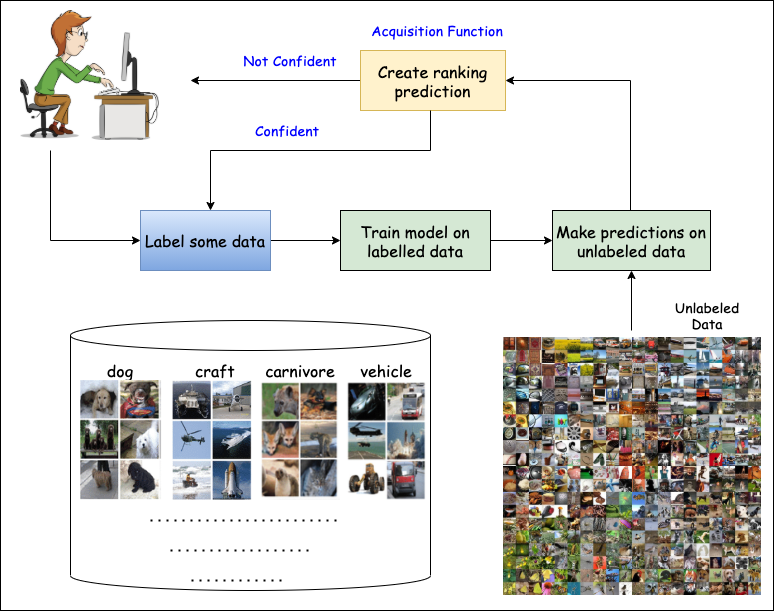}
	  \caption{Representation of the deep active learning process}\label{fig2}
\end{figure*}

\section{Active learning and Acquisition Functions}
Figure 1 shows how traditional or passive supervised learning works. Examples in an unlabeled dataset are labeled by one or more human "experts", and the labeled dataset is used to train the classifier. Modern deep learning classifiers need large labeled datasets, and acquiring such datasets with human involvement is expensive and error-prone. Active learning attempts to reduce the cost of labeling datasets by starting with a small human-labeled dataset and then labeling a large unlabeled dataset with minimal human involvement. The classifier trained with the small labeled dataset is used to classify the unlabeled examples, and those examples that the classifier is least confident about are sent to an expert for classification. The expert or the oracle hand classifiers these examples. These examples are incorporated into the dataset to create an extended dataset and the process is repeated as many times as necessary. The objective in active learning is to minimizing the number of examples that need to be hand-classified by oracle to achieve very high classifier accuracy at the end of the iterations.

Acquisition functions are used to select examples classified by current classifier that are sent to the oracle to label by hand. The approach we choose to use is based on computation of uncertainty in classification. Only the examples with the highest amount of uncertainty are sent to the oracle. We experiment with three acquisition functions in this paper.

\subsection{Least Confidence Based Uncertainty}
As shown in Equation (\ref{Equation1}), the Least Confidence measure computes the difference between 100\% confidence and the most confident prediction to compute the sample probability of querying a label for example \cite{settles2009active}:\\
\begin{equation}
\label{Equation1}
x^*_{lc}=\underset{x}{\mathrm{argmax}}\,  (1-P_\theta(\hat{y} | x ))
\end{equation}
where $x$ is a sample, $P_\theta(\hat{y} | x )$ denotes the probability of acquisition associated with the most probable class, $\theta$ represents parameters of a model, and $\hat{y}=\underset{y}{\mathrm{argmax}}\, (y | x )$. \\

\subsection{Margin Sampling Based Uncertainty}
Margin sampling, defined in Equation (\ref{Equation2}), estimates the uncertainty in classification of a sample by computing the difference in the confidence of the top two predictions \cite{settles2009active}. The sampling probability of querying a label can be written as:\\
\begin{equation}
\label{Equation2}
x^*_{M}=\underset{x}{\mathrm{argmax}}\,  (P_\theta(\hat{y_1} | x )-P_\theta(\hat{y_2} | x )) 
\end{equation}
where probability of $P_\theta(\hat{y_1} | x )$ is the probability of the highest ranked class, $P_\theta(\hat{y_2} | x )$ is the second ranked class, and $\hat{y_1}$ and $\hat{y_2}$ correspond to the two top most predictions for $x$ given the model parameters $\theta$.

\subsection{Entropy Based Uncertainty}
As expressed in Equation (\ref{Equation3}), the entropy-based uncertainty is an information theory based method to estimate uncertainty \cite{shannon2001mathematical}. The sampling probability of querying a label can be written as:\\
\begin{equation}
\label{Equation3}
x^*_{H}=\underset{x}{\mathrm{argmax}}\,  -\sum_{i}^{} P_\theta(\hat{y_i} | x )\log P_\theta(\hat{y_i} | x ) 
\end{equation}

where $x$ is the sample to be labeled, $\hat{y_i}$ is the $i$-th possible label, and $P(\hat{y_i}|x)$ is the predicted probability of the $i$-th label given the input $x$. The uncertainty score of a sample is the sum of the entropy of its predicted probabilities for all possible labels. The sample with the highest uncertainty score is selected for labeling. The intuition behind this formula is that the model is uncertain about the correct label for a sample when the predicted probabilities for all possible labels are close to each other, resulting in a high entropy score. Therefore, selecting the samples with the highest entropy score helps the model to learn more effectively by focusing on the most informative samples.
\section{Methodology}
In this paper, we develop a number of approaches for active learning and compare them. The approaches we develop can be categorized into two groups: \romannumeral 1) SVM combined with CNN-based deep neural networks, and \romannumeral 2) Transformer-based deep neural networks. SVMs are excellent linear classifier that consistently produce high accuracy for datasets given in term of clearly identified features such as age, income, height, weight, and blood pressure, that are high-level and deemed important for the task at hand. Our hypothesis is that if high-level abstract features are first extracted by a CNN-based deep learning model from "raw" features, these high level features are likely to produce excellent results in malware classification, if malware are represented as images.

The second set of models with which we experiment is motivated by the hypothesis that the Vision Transformer (ViT) \cite{dosovitskiy2020image} that has recently achieved excellent results in many vision-related tasks, including classification, is likely to achieve good results with malware classification as well. Since a good classifier is likely to perform well in active learning, we use these two categories of classification in our active learning experiments. 

\subsection{Support Vector Machine}

The binary Support Vector Machine (SVM) derives a hyperplane that maximizes the separating margin between the positive and negative classes \cite{noble2006support}. Suppose we have two classes of malicious files of $m$ data points ${\{{\textbf{x}_i},{y_i}\}^{m}_{i=1}}$ with $n$-dimensional features ${\textbf{x}_i \in \mathbb{R}^n}$ and labels $y_i \in \{+1,-1\}$. The linear separator in an SVM is the hyperplane $\textbf{w}^T \textbf{x} + b = 0$ that separates the data points to two classes with a classification rule based on sign($\textbf{w}^T \textbf{x} + b$). A positive example \textbf{x} is in +1 class otherwise -1 class. 

A limitation of the linear SVM is that the hyperplane cannot separate most datasets in the real world. A workaround is to apply a non-linear transformation to the data points before applying SVM. We can achieve the desired effect of getting a non-linear decision boundary without changing how the SVM works internally. \\

The first set of the models we use for active learning consists of the use a CNN-based deep learning architecture such as Inception-V4 \cite{szegedy2017inception} to extract high-level latent or abstract features, and then classify the features using an SVM. Before passing to an SVM, we use Principal Component Analysis (PCA) \cite{turk1991eigenfaces} to select relevant features to make the classification by the SVM more efficient. Inception-V4 is considered one of the best transfer learning models for feature extraction \cite{eltehewy2023efficient} due to its architecture that includes both inception and residual blocks, which allows for efficient use of computational resources and improved accuracy on image classification tasks. Additionally, Inception-V4 has a large number of layers, which enables it to capture more complex and abstract features from images compared to other transfer learning models \cite{tan2018survey,niu2020decade,chouhan2020novel}. In Principal Component Analysis (PCA), a reduced set of feature values are obtained by projecting the original data onto a lower-dimensional space defined by the principal components. This is done by calculating the dot product between the original data and the principal component vectors, resulting in a new set of values for each example that represents their position along each principal component axis. These new values are the composite feature values and are used to create the reduced-dimensionality dataset for further analysis.

We propose an Active Learning model that uses Inception-V4+PCA with different variations of SVM models to perform classification with malware image data as we also prepare to identify data examples that the classifier is most uncertain about. We used SVM models and employ the entropy-based approach to compute the amount of uncertainty for each algorithm; see Figure 3.\\

SVM models are specifically designed to handle uncertinty in classification of data examples, which a straightforward SVM model cannot do effectively. Uncertainty in classification of data can arise from various sources such as missing values, measurement errors, noise, and ambiguity in data interpretation. These models incorporate uncertainty information into their learning process, which enables them to make more accurate predictions and enhance their overall performance. The incorporation of uncertainty information is a critical factor in making these models more robust and reliable. Therefore, these models have a significant advantage over the standard SVM models in scenarios where uncertainty needs to be considered.

The \textbf{Uncertainty-aware Twin Support Vector Machines (UTSVM)}, is a variant of Support Vector Machines (SVMs) that takes into account the uncertainty in the training data. The idea is to model the sample uncertainty as a Gaussian distribution with mean equal to the sample and covariance matrix representing the uncertainty in the sample. The goal is to learn two classification functions, $f_1(\mathbf{x})$ and $f_2(\mathbf{x})$, that maximize the margin between the two classes, while also minimizing the uncertainty in the data.\\

The \textbf{Power SVM (PSVM)} handles the uncertainty in classifying an example by introducing a new loss function that penalizes the classification error by a weight that is proportional to the degree of uncertainty of the example. Specifically, the uncertainty is measured as the distance between the example and its nearest neighbors in the feature space. This weight is then used to adjust the contribution of the exemplar to the SVM objective function. The framework allows the uncertainty to be taken into account in the learning process and can lead to improved generalization performance.

In the \textbf{SVM with Gaussian Sample Uncertainty (SVM-GSU)} method, the certainty of the input data is also modeled by assigning weights to each data point based on the estimated uncertainty. The weights are incorporated into the objective function of the linear maximum margin classifier to maximize the margin between the classes while also minimizing the classification error. Specifically, the contribution of each data point to the objective function is re-weight such that uncertain points are given lower weight and more certain points are given higher weight. This allows the classifier to focus more on the more certain data points while still incorporating the uncertain ones in the learning process.

The \textbf{Twin Bounded Support Vector Machines (TBSVM)} is an improved version of Twin Support Vector Machines (TWSVM) \cite{mangasarian2005multisurface}. The algorithm incorporates uncertainty information into the objective function to make it more robust to noisy and uncertain data. Specifically, the uncertainty information is used to adjust the weight of the penalty term in the objective function, which controls the trade-off between margin maximization and error minimization. 
\begin{figure} [!ht]
\centering
	\includegraphics[width=10cm, height=19cm, frame]{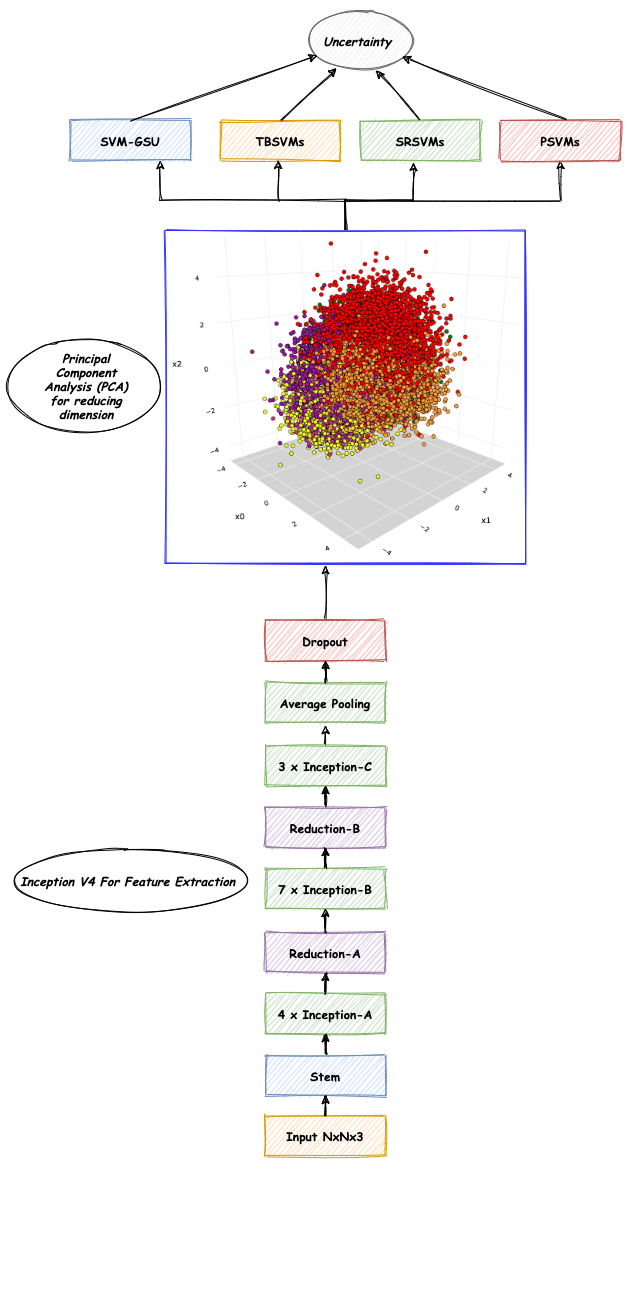}
	\caption{Proposed Inception-V4+PCA with different SVMs algorithms}\label{fig3}
\end{figure}

\subsection{Vision Transformer}
Vision Transformer is a new state-of-the-art attention based model for image classification \cite{carion2020end,touvron2019fixing,yuan2021tokens}. Its accuracy has surpassed the best convolutional neural networks for image classification \cite{chen2021crossvit,bazi2021vision,fan2021multiscale,bhojanapalli2021understanding}. We propose a novel classification approach to malware and compute uncertainties in malware classification using the Vision Transformer architecture. To use ViT, we need to obtain patches from the input. The Vision Transformer takes the input image and splits it into grid-based patches of 16 by 16 pixels. These patches are then flattened by a linear transformation matrix to create vectors. All patches go through a linear projection layer to produce vectors $z^1_1$ to $z^9_1$ as shown in Figure 4. These vectors are arranged in sequence from top to bottom. 

The vector for each patch also gets a positional embedding, and the Transformer is used to predict what the image might represent. The Transformer is unlike the CNN because it has the freedom to look everywhere at the image from the start, as shown in Figure 5.
\\
\begin{figure*}[!ht]
\centering
	\includegraphics[width=13cm, height=5cm, frame]{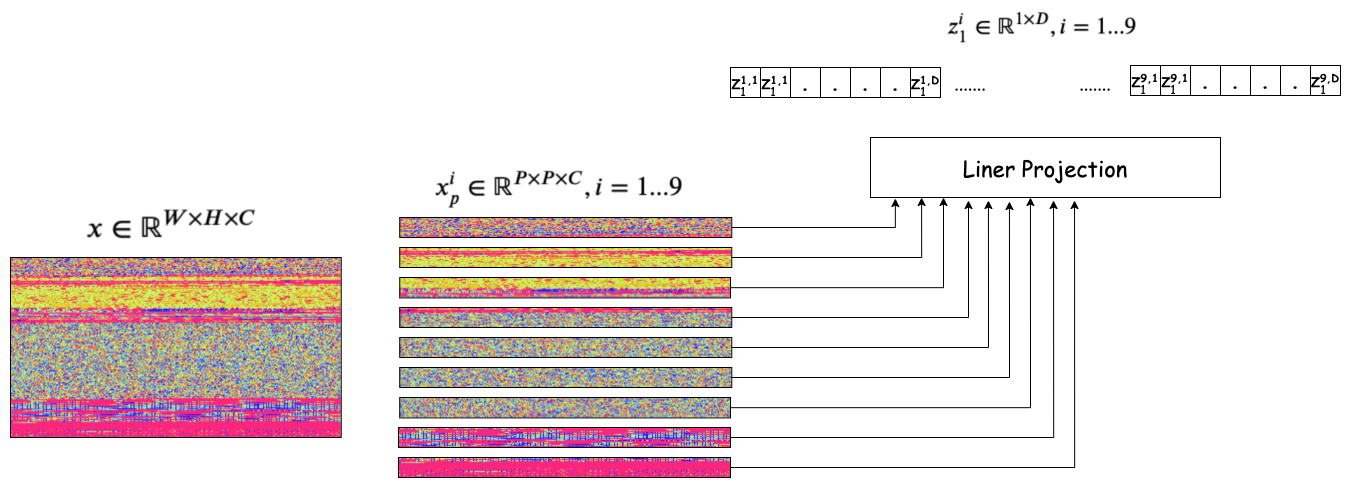}
	\caption{patches to features}
\end{figure*}

The linear projection can be performed using a dense or linear layer without the bias term as shown in Figure 4. 
\begin{figure} [!ht]
\centering
\includegraphics[width=4cm, height=3cm, frame]{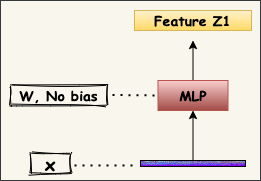}
	\caption{Linear projection}
\end{figure} 

The next step is to add position embedding. Position embedding adds a representation of the unique position to the linear projection of each patch, and the Vision Transformer knows the arrangement of the sequences during training. ViT assigns a position embedding to learnable class embeddings. The final layer feature vector corresponding to this learnable class embedding is used by the MLP head for classification. The most important part of Vision Transformer is the encoder. Transformer encoder is made of multiple encoder modules with identical network structure. This is done by stacking multiple encoder modules together; see Figure 5. 

The transformer network is unable to compute the confidence in the predictions, which also makes it difficulties to estimate the uncertainty (or confidence) in classification. Thus, combining the transformer network with Bayesian theory can handle uncertainty in classification. Bayesian neural networks can estimate uncertainty in changes of weight and bias parameters of the neural networks using probability distributions \cite{kendall2017uncertainties,bensaoud2025optimized}.

In the neural network overfitting occurs when we choose network weights that do satisfactorily in training but poorly on testing objective function. Therefore, we remove the MLP head and add a BNN \cite{blundell2015weight} with Gaussian Distribution to capture uncertainty in the predictions. BNNs solve the overfitting problem by modeling uncertainty in the weights and using a stochastic variational inference algorithm. We apply Gaussian distribution in BNN to approximate the correct posterior and to improve the BNN in handling uncertainty. For setting up the Bayesian network , we modify Gaussian distribution by using different hyperparameters values, 0.1, 0.3, and 0.5, to improve the performance of the model. Our approach is to let a classifier not only predict the malware on a given sample but also present the amount of uncertainty of the predicted label.
\begin{figure*}[!ht]
\centering
	\includegraphics[width=16cm, height=17cm, frame]{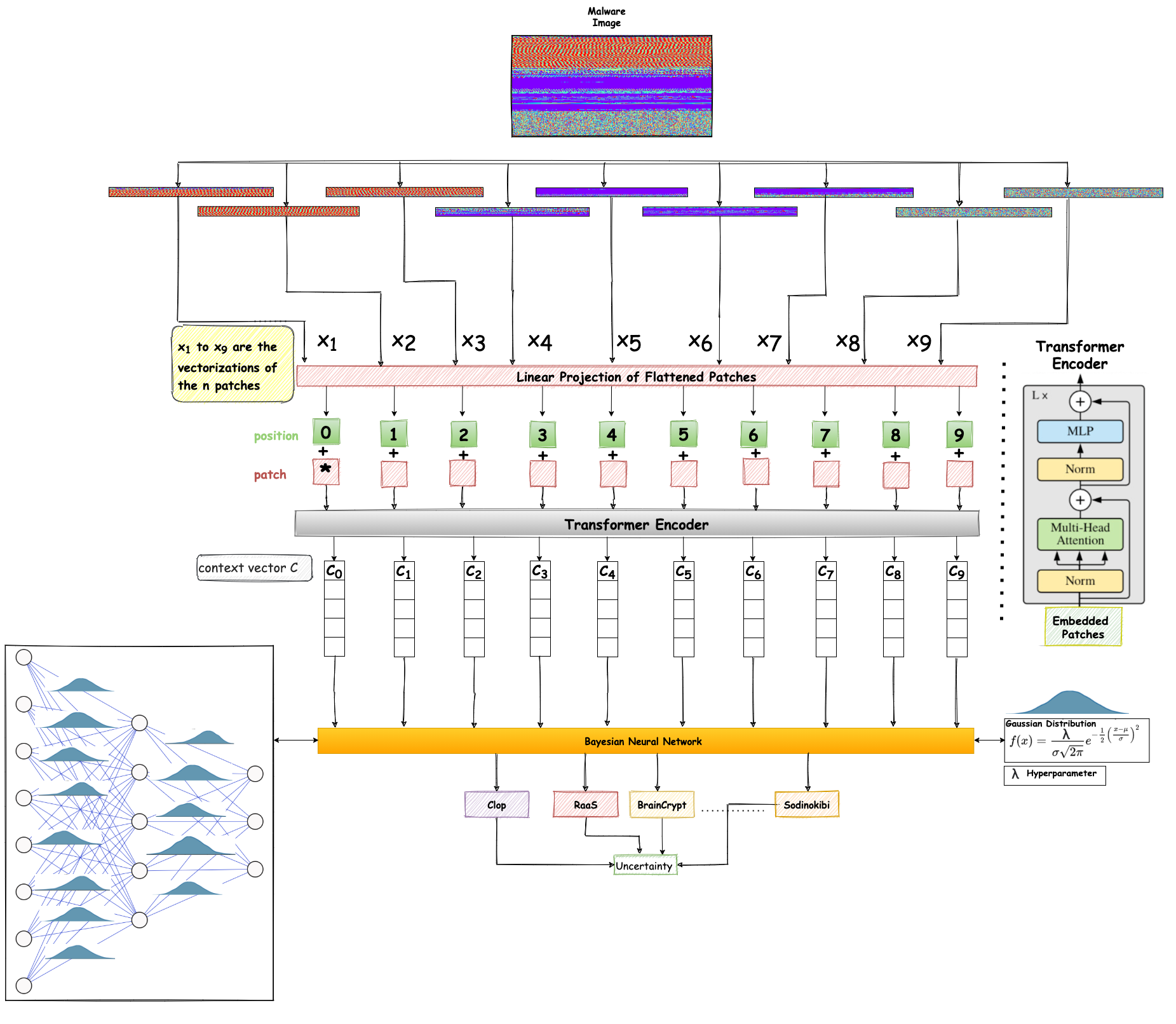}
	\caption{Proposed Vision Transformer with Bayesian Neural Network }
\end{figure*}

\subsection{Handling Uncertainty with Bayesian Neural Networks}
BNNs provide a principled way to incorporate and handle uncertainty in neural network models by treating the network's weights as random variables and placing a probability distribution over them. Unlike traditional neural networks, which produce point estimates of weights through optimization, BNNs maintain a distribution over the weights, thus allowing for probabilistic predictions.

\subsubsection{Model Formulation}

Consider a BNN with \( L \) hidden layers. Let \( \mathbf{x} \in \mathbb{R}^d \) be the input and \( \mathbf{y} \in \mathbb{R}^K \) be the output. The network is parameterized by weights \( \mathbf{W}^{(l)} \) and biases \( \mathbf{b}^{(l)} \) for \( l = 1, \ldots, L \). The output at layer \( l \) is denoted as \( \mathbf{z}^{(l)} \), with the input layer being \( \mathbf{z}^{(0)} = \mathbf{x} \).

In Equation (\ref{Equation4}), the transformation at each layer \( l \) is defined as:
\begin{equation}
\label{Equation4}
\mathbf{z}^{(l)} = \phi_l (\mathbf{W}^{(l)} \mathbf{z}^{(l-1)} + \mathbf{b}^{(l)}),
\end{equation}
where \( \phi_l \) is the non-linearity applied element-wise.  The final output is given by Equation (\ref{Equation5}):
\begin{equation}
\label{Equation5}
\mathbf{y} = \mathbf{z}^{(L)}.
\end{equation}

\subsubsection{Likelihood and Prior}

The likelihood of the observed data \( D = \{(\mathbf{x}_i, \mathbf{y}_i)\}_{i=1}^N \) given the parameters \( \omega \) (all weights and biases) is given by Equation (\ref{Equation6}):
\begin{equation}
\label{Equation6}
p(\mathbf{y} | \mathbf{x}, \omega).
\end{equation}
We place a Gaussian prior over the parameters, as shown in Equation (\ref{Equation7}):
\begin{equation}
\label{Equation7}
p(\omega) = \prod_{l=1}^L \mathcal{N}(\mathbf{W}^{(l)} | \mathbf{0}, \sigma^2 \mathbf{I}) \mathcal{N}(\mathbf{b}^{(l)} | \mathbf{0}, \sigma^2 \mathbf{I}),
\end{equation}
where \( \sigma \) is a hyperparameter that controls the variance of the prior.

\subsubsection{Variational Inference with Gaussian Approximations}

To approximate the posterior distribution \( p(\omega | D) \), we use variational inference with a Gaussian variational family. Specifically, we approximate the posterior with a Gaussian distribution, as shown in Equation (\ref{Equation8}):

\begin{equation}
\label{Equation8}
q_{\theta}(\omega) = \mathcal{N}(\omega | \mu, \Sigma),
\end{equation}
where \( \theta = \{\mu, \Sigma\} \) are the parameters of the variational distribution, representing the mean vector and the covariance matrix, respectively.

\subsubsection{Variational Objective}

The objective of variational inference is to minimize the Kullback-Leibler (KL) divergence between the variational distribution and the true posterior, as given in Equation (\ref{Equation9}):

\begin{equation}
\label{Equation9}
\theta^* = \underset{\theta \in \Theta}{\mathrm{argmin}} \, \mathrm{KL} \left( q_{\theta}(\omega) \, || \, p(\omega | D) \right).
\end{equation}
This is equivalent to maximizing the evidence lower bound (ELBO), as shown in Equation (\ref{Equation10}):
\begin{equation}
\label{Equation10}
\mathcal{L}_{\text{VI}}(\theta) = \mathbb{E}_{q_{\theta}(\omega)} [\log p(D | \omega)] - \mathrm{KL}(q_{\theta}(\omega) \, || \, p(\omega)).
\end{equation}

The Kullback-Leibler (KL) divergence term for Gaussian distributions, as expressed in Equation (\ref{kl_gaussian}), is formulated as follows:
\begin{fleqn}[\parindent]
\begin{equation}
\label{kl_gaussian}
\begin{split}
 \mathrm{KL}\left(\mathcal{N}(\mu, \Sigma) \, || \, \mathcal{N}(0, \sigma^2 \mathbf{I})\right) = \\
\frac{1}{2} \left( \mathrm{tr}(\Sigma^{-1} \Sigma) + \mu^T \Sigma^{-1} \mu - K + \log \frac{\det \sigma^2 \mathbf{I}}{\det \Sigma} \right)
\end{split}
\end{equation}
\end{fleqn}

\subsubsection{Variational Predictive Distribution}

The variational predictive distribution provides an approximation to the true predictive distribution \( p(\mathbf{y}^* | \mathbf{x}^*, D) \) for a new input \( \mathbf{x}^* \), as expressed in Equation (\ref{eq:predictive_distribution}):
\begin{equation}
\label{eq:predictive_distribution}
p(\mathbf{y}^* | \mathbf{x}^*, D) \approx \int p(\mathbf{y}^* | \mathbf{x}^*, \omega) q_{\theta^*}(\omega) d\omega.
\end{equation}

Given the Gaussian approximation, the predictive distribution can be approximated using Monte Carlo sampling, as shown in Equation (\ref{eq:monte_carlo_sampling}):
\begin{equation}
\label{eq:monte_carlo_sampling}
p(\mathbf{y}^* | \mathbf{x}^*, D) \approx \frac{1}{M} \sum_{m=1}^M p(\mathbf{y}^* | \mathbf{x}^*, \omega_m),
\end{equation}
where \( \omega_m \sim q_{\theta^*}(\omega) \).

\subsubsection{Quantifying Predictive Uncertainty}

\begin{itemize}
    \item \textbf{Entropy-based Uncertainty}: The entropy of the predictive distribution, which quantifies the uncertainty in predictions, is expressed in Equation (\ref{eq:entropy}):
\begin{equation}
\label{eq:entropy}
\mathbb{H}[\mathbf{y}^* | \mathbf{x}^*, D] = -\int p(\mathbf{y}^* | \mathbf{x}^*, D) \log p(\mathbf{y}^* | \mathbf{x}^*, D) d\mathbf{y}^*.
\end{equation}
High entropy indicates high uncertainty, while low entropy indicates confident predictions.

    \item \textbf{Moment-based Uncertainty}: The variance of the predictive distribution, which serves as an additional metric for uncertainty, is detailed in Equation (\ref{eq:variance}):
\begin{equation}
\label{eq:variance}
\mathrm{Var}[\mathbf{y}^* | \mathbf{x}^*, D] = \mathbb{E}[\mathbf{y}^{*2} | \mathbf{x}^*, D] - (\mathbb{E}[\mathbf{y}^* | \mathbf{x}^*, D])^2,
\end{equation}
where the expectation is defined in Equation (\ref{eq:expectation}):
\begin{equation}
\label{eq:expectation}
\mathbb{E}[\mathbf{y}^* | \mathbf{x}^*, D] = \int \mathbf{y}^* p(\mathbf{y}^* | \mathbf{x}^*, D) d\mathbf{y}^*.
\end{equation}

\end{itemize}

\subsection{New Proposal with Hypotheses}

We propose a novel method for Bayesian Neural Networks that enhances the handling of uncertainty by combining advanced variational inference techniques with improved prior distributions. This method aims to provide more accurate uncertainty quantification and robust predictions in neural networks.

\subsubsection{Hypotheses}
To enhance the performance of Bayesian neural networks integrated with Vision Transformers, we propose two primary hypotheses. These hypotheses are centered around improving the variational approximation and refining the prior distributions.

\begin{itemize}
    \item \textbf{Improved Variational Approximation}: The new variational distribution \( q_{\theta}(\omega) \) will better approximate the true posterior \( p(\omega | D) \) by minimizing the KL divergence more effectively using advanced optimization techniques and a richer family of distributions.

    \item \textbf{Enhanced Prior Distributions}: Introducing structured priors \( p(\omega) \) based on domain knowledge, hierarchical Bayesian models will lead to more informative and regularized posterior distributions.
\end{itemize}

\subsubsection{Mathematical Formulation}

The variational inference loss function, denoted as \( \mathcal{L}_{\text{VI}} \), is formally described in Equation (\ref{eq:vi_loss}):
\begin{equation}
\label{eq:vi_loss}
\mathcal{L}_{\text{VI}}(\theta) = \mathrm{KL} \left( q_{\theta}(\omega) \, || \, p(\omega) \right) - \mathbb{E}_{q_{\theta}(\omega)} [\log p(D | \omega)].
\end{equation}

\textbf{Hypothesis 1} posits that the optimal variational distribution, \( q_{\theta^*}(\omega) \), is determined as outlined in Equation (\ref{eq:optimal_variational}):
\begin{equation}
\label{eq:optimal_variational}
\theta^* = \underset{\theta \in \Theta}{\mathrm{argmin}} \, \mathcal{L}_{\text{VI}}(\theta).
\end{equation}

\textbf{Hypothesis 2} suggests that the structured prior \( p(\omega) \), defined in Equation (\ref{eq:structured_prior}), enhances posterior regularization:
\begin{equation}
\label{eq:structured_prior}
p(\omega) = \prod_{l=1}^L \mathcal{N}(\mathbf{W}^{(l)} | \mathbf{0}, \sigma^2 \mathbf{I}) \mathcal{N}(\mathbf{b}^{(l)} | \mathbf{0}, \sigma^2 \mathbf{I}),
\end{equation}
where \( \sigma \) is chosen based on domain-specific insights.

The new variational predictive distribution, expressed in Equation (\ref{eq:variational_predictive}), approximates the true predictive distribution:
\begin{equation}
\label{eq:variational_predictive}
p(\mathbf{y}^* | \mathbf{x}^*, D) \approx \int p(\mathbf{y}^* | \mathbf{x}^*, \omega) q_{\theta^*}(\omega) d\omega.
\end{equation}

\textbf{Predictive Uncertainty}:\\
To effectively quantify the uncertainty in predictions, we employ two complementary methods: entropy-based and moment-based uncertainty, which provide detailed insights into the model's confidence levels.

\begin{itemize}
    \item \textbf{Entropy-based}: The uncertainty is evaluated through the entropy of the predictive distribution, as defined in Equation~(\ref{eq:entropy}):
    \begin{equation}
    \label{eq:entropy}
    \mathbb{H}[\mathbf{y}^* | \mathbf{x}^*, D] = -\int p(\mathbf{y}^* | \mathbf{x}^*, D) \log p(\mathbf{y}^* | \mathbf{x}^*, D) d\mathbf{y}^*.
    \end{equation}

    \item \textbf{Moment-based}: The variance of the predictive distribution, described in Equation~(\ref{eq:variance}), offers an alternative perspective by capturing the spread of predictions:
    \begin{equation}
    \label{eq:variance}
    \mathrm{Var}[\mathbf{y}^* | \mathbf{x}^*, D] = \mathbb{E}[\mathbf{y}^{*2} | \mathbf{x}^*, D] - (\mathbb{E}[\mathbf{y}^* | \mathbf{x}^*, D])^2.
    \end{equation}
\end{itemize}

\subsubsection{Gaussian Distribution with $\lambda$ in Deep Bayesian Neural Networks}

We applied a Gaussian distribution to model uncertainty in the parameters of a Deep Bayesian Neural Network (DBNN). The network parameters, denoted by \( \omega \), include weights and biases. The likelihood of the observed data given the parameters \( \omega \) is expressed as \( p(\mathbf{y} | \mathbf{x}, \omega) \), where \( \mathbf{y} \) represents the output and \( \mathbf{x} \) the input.

To incorporate uncertainty into the parameters, we assumed a Gaussian prior distribution over the weights and biases, as formulated in Equation~(\ref{eq:gaussian_prior}):
\begin{equation}
\label{eq:gaussian_prior}
p(\omega) = \prod_{i=1}^{N} \mathcal{N}(\omega_i | \mu_i, \sigma_i^2),
\end{equation}
where \( \mu_i \) and \( \sigma_i^2 \) denote the mean and variance of the Gaussian distribution for the \( i\)-th parameter, respectively.

During training, our objective was to determine the posterior distribution \( p(\omega | \mathbf{D}) \), where \( \mathbf{D} \) denotes the training dataset. According to Bayes' theorem, the posterior is proportional to the product of the likelihood and the prior, as described in Equation~(\ref{eq:posterior_distribution}): 
\begin{equation}
\label{eq:posterior_distribution}
p(\omega | \mathbf{D}) \propto p(\mathbf{D} | \omega) \cdot p(\omega),
\end{equation}
where \( p(\mathbf{D} | \omega) \) represents the likelihood function of the data given the parameters.

In practice, computing the exact posterior distribution is often intractable, so we approximated it using variational inference. We introduced a variational distribution \( q(\omega | \theta) \), parameterized by \( \theta \), to approximate the true posterior \( p(\omega | \mathbf{D}) \). The objective was to minimize the KL divergence between the variational distribution and the true posterior, formulated in Equation~(\ref{eq:kl_variational}):

\begin{equation}
\label{eq:kl_variational}
\theta^* = \arg\min_{\theta} \, \text{KL}(q(\omega | \theta) \parallel p(\omega | \mathbf{D})).
\end{equation}

The predictive distribution for a new input \( \mathbf{x}^* \) can be obtained by integrating over the posterior distribution, as expressed in Equation~(\ref{eq:predictive_distribution}):

\begin{equation}
\label{eq:predictive_distribution}
p(\mathbf{y}^* | \mathbf{x}^*, \mathbf{D}) = \int p(\mathbf{y}^* | \mathbf{x}^*, \omega) \, p(\omega | \mathbf{D}) \, d\omega
\end{equation}

This integration is typically approximated using Monte Carlo sampling from the variational distribution.

Additionally, we enhanced the predictive distribution \( p(\mathbf{y} | \mathbf{x}, \omega) \) by applying a Gaussian distribution with a scaling parameter \( \lambda \) , as shown in Equation~(\ref{eq:enhanced_predictive_distribution}):

\begin{equation}
\label{eq:enhanced_predictive_distribution}
f(x) = \lambda \frac{1}{\sigma \sqrt{2\pi}} e^{-\frac{1}{2}\left(\frac{x-\mu}{\sigma}\right)^2}.
\end{equation}

In this formulation, \( \lambda \) controls the scaling of the distribution, allowing us to fine-tune the uncertainty modeling further. This modification enabled our model to better capture the uncertainties in the data.

\subsection{High-Value Sample Selection in Active Learning}

The selection of high-value samples is crucial for enhancing the efficiency and effectiveness of active learning models. In our study, \textbf{high-value samples} are defined as instances within the dataset expected to significantly impact model performance and generalization, identified based on criteria such as rarity, feature complexity, and task relevance. The rationale for selecting these samples is to maximize labeling efficiency and enhance dataset quality by prioritizing instances likely to challenge the model or reveal potential weaknesses. This approach facilitates more robust training while aiding in uncovering critical patterns and anomalies.

While uncertainty is an important criterion for determining the value of samples, it is not the sole factor. Incorporating multiple criteria into the selection process can lead to more effective active learning. Factors such as sample diversity, representativeness of the data, and proximity to decision boundaries contribute to identifying high-value samples. SVMs excel at pinpointing uncertain samples near the decision boundary, while ViTs leverage attention mechanisms to focus on crucial input regions. BNNs enhance uncertainty quantification, identifying samples with high predictive uncertainty. However, a comprehensive sample selection strategy requires combining these approaches with additional criteria, such as information density. By incorporating active learning principles, our methodology optimizes the selection of \textbf{high-value samples} through an iterative process where the model selectively queries the most informative data points for labeling, continuously refining the dataset and concentrating on examples that meaningfully contribute to the model's learning trajectory.

\section{Experiments}
In this section, we present a set of experiments to evaluate two proposed models. The training and testing set contain 90,000 and 10,000 labeled images respectively. We consider the training set as the initial unlabeled pool $U_{1000,000}$ and we divided the $U_{1000,000}$ into $U_{100,000}$ each cycle for all experiments.

For dataset, we collected a dataset of 1,000,000 malware samples from sources like VirusTotal\footnote{\url{https://www.virustotal.com}} , Contagio\footnote{\url{http://contagiodump.blogspot.com}}, malshare\footnote{\url{https://www. malshare.com}} and VirusShare \footnote{\url{https://www. virusshare.com}} . Based on these samples, we generated an extensive collection of color images representing diverse malware executable files for Windows, Android, Linux, MacOS, and iOS operating systems.

\subsection{Hypothesis 1: Improved Variational Approximation}

To evaluate Hypothesis 1, we conducted experiments comparing the performance of our proposed variational inference method with traditional techniques in BNNs. The key aspect of this hypothesis is that our method aims to provide a more accurate approximation of the posterior distribution through the use of Gaussian variational approximation. This was achieved by refining the variational family to better capture the complexities of the true posterior.

The experiments involved training BNNs on the malware image dataset and assessing their predictive performance and uncertainty quantification. Specifically, we measured the number of samples classified as "not confident" across multiple training cycles. Our results demonstrated that the improved variational approximation significantly reduces the number of uncertain samples, indicating better model confidence and reliability. The reduction in uncertainty is evident from the consistent decrease in the number of not confident samples across cycles.

\subsection{Hypothesis 2: Enhanced Prior Distributions}

Hypothesis 2 explores the impact of enhanced prior distributions on the performance of BNNs. Traditional BNNs often rely on simple priors, such as Gaussian distributions with fixed parameters. In this hypothesis, we propose using more informative and flexible priors to better reflect the underlying distribution of the data.

We experimented with various prior configurations and analyzed their effect on model performance and uncertainty handling. The enhanced priors were designed to incorporate domain-specific knowledge and empirical data characteristics. As with Hypothesis 1, we evaluated the models using the same malware image dataset, tracking the number of not confident samples over multiple training cycles.

The results indicate that enhanced prior distributions contribute to more robust uncertainty quantification and improved predictive performance. The models with enhanced priors showed a marked decrease in the number of uncertain samples, similar to the results observed with improved variational approximation. This demonstrates that incorporating more informative priors can lead to more confident and reliable predictions.

Overall, our results support the effectiveness of utilizing Bayesian Neural Networks with Gaussian variational approximation and enhanced prior distributions for handling uncertainty in various machine learning tasks.

\begin{figure*}[!ht]
\centering
\includegraphics[width=0.8\textwidth]{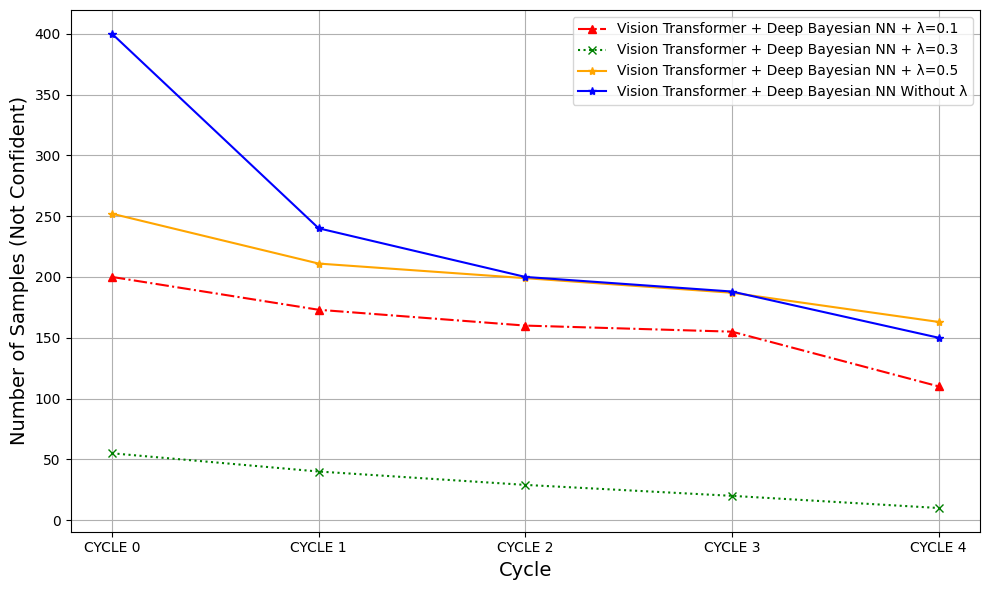}
\caption{Effect of $\lambda$ on Uncertainty Handling }
\label{fig:lambda_effect}
\end{figure*}
Figure 7 visualizes the impact of varying the value of $\lambda$ on the number of samples classified as "not confident" by the Vision Transformer + DBNN across different training cycles. The plot compares four scenarios: without $\lambda$ (blue line with circular markers), $\lambda = 0.1$ (green line with square markers), $\lambda = 0.3$ (red line with triangular markers), and $\lambda = 0.5$ (purple line with cross markers). The x-axis represents the training cycles (from CYCLE 0 to CYCLE 4), and the y-axis indicates the number of samples not confidently classified. The results demonstrate that increasing $\lambda$ generally reduces the number of uncertain samples, with $\lambda = 0.3$ showing the most significant reduction, while $\lambda = 0.5$ exhibits a higher number of uncertain samples compared to $\lambda = 0.3$. This suggests that an optimal $\lambda$ value can enhance the model's confidence in its predictions, improving uncertainty handling. The plot includes labels for clarity and a legend to distinguish between different $\lambda$ values, emphasizing the overall trend and effectiveness of $\lambda$ in reducing model uncertainty.

\begin{figure*}[!ht]
\centering
\includegraphics[width=0.8\textwidth]{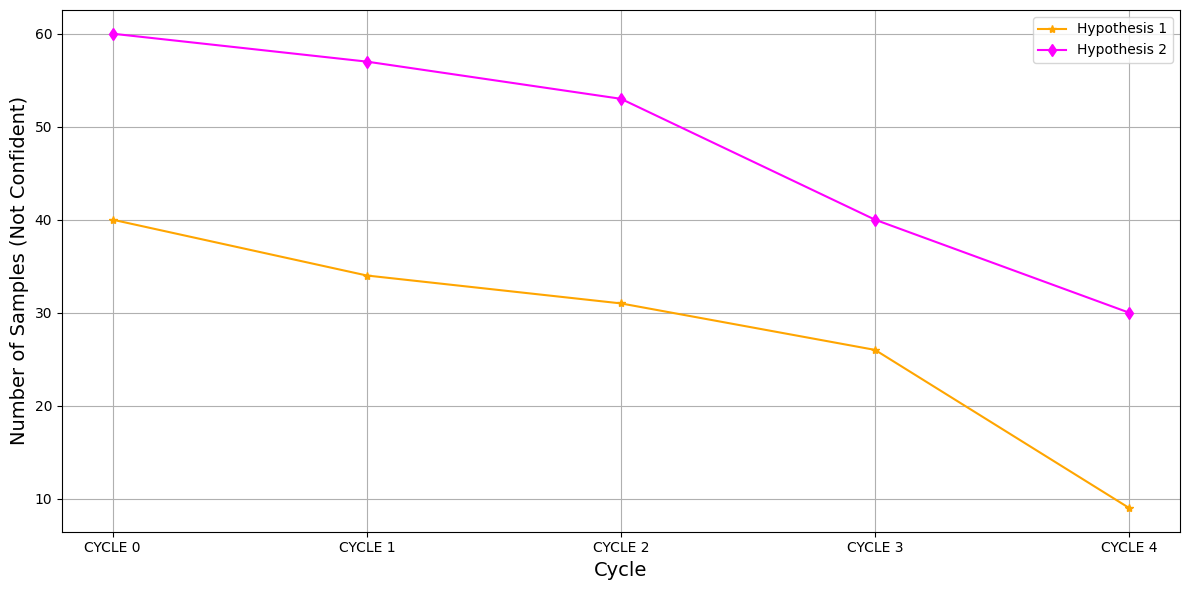}
\caption{Comparison of Uncertainty Handling Between Hypothesis 1 and Hypothesis 2}
\label{fig:uncertainty_handling_comparison}
\end{figure*} 
Figure 8 presents a comparison of the number of samples classified as "not confident" by the models under Hypothesis 1 and Hypothesis 2 across five training cycles (CYCLE 0 to CYCLE 4). Hypothesis 1 (orange line with star markers) consistently results in fewer uncertain samples compared to Hypothesis 2 (magenta line with diamond markers). This indicates that Hypothesis 1 is more effective in reducing uncertainty throughout the training process. The x-axis represents the training cycles, while the y-axis indicates the number of samples not confidently classified. The plot includes labels for clarity and a legend to differentiate between the two hypotheses, highlighting the overall trend and the effectiveness of each hypothesis in managing uncertainty.
\begin{figure*}[!ht]
\centering
\includegraphics[width=0.8\textwidth]{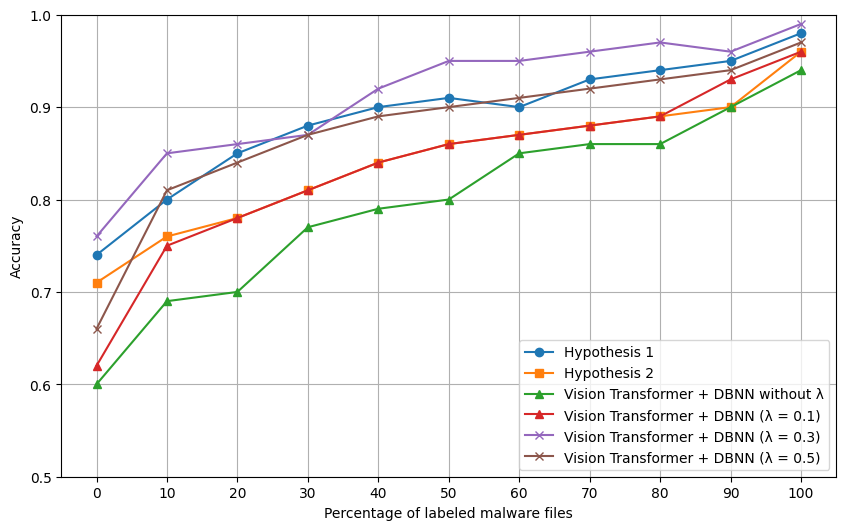}
\caption{Impact of Labeled Malware Data Percentage on Model Accuracy}
\end{figure*}
Figure 9 above illustrates the impact of increasing the percentage of labeled malware files on the accuracy of different models in malware classification. Each line represents a different model configuration, including Hypothesis 1, Hypothesis 2, Vision Transformer combined with a Bayesian Neural Network without $\lambda$, and Vision Transformer combined with a Bayesian Neural Network with various $\lambda$ values. As the percentage of labeled malware files increases, the accuracy of all models generally improves. However, it's evident that Vision Transformer combined with a Bayesian Neural Network with $\lambda$=0.3 consistently outperforms other configurations across different percentages of labeled data, demonstrating its effectiveness in handling malware classification tasks.
\begin{figure*}[!ht]
\centering
\includegraphics[width=0.8\textwidth]{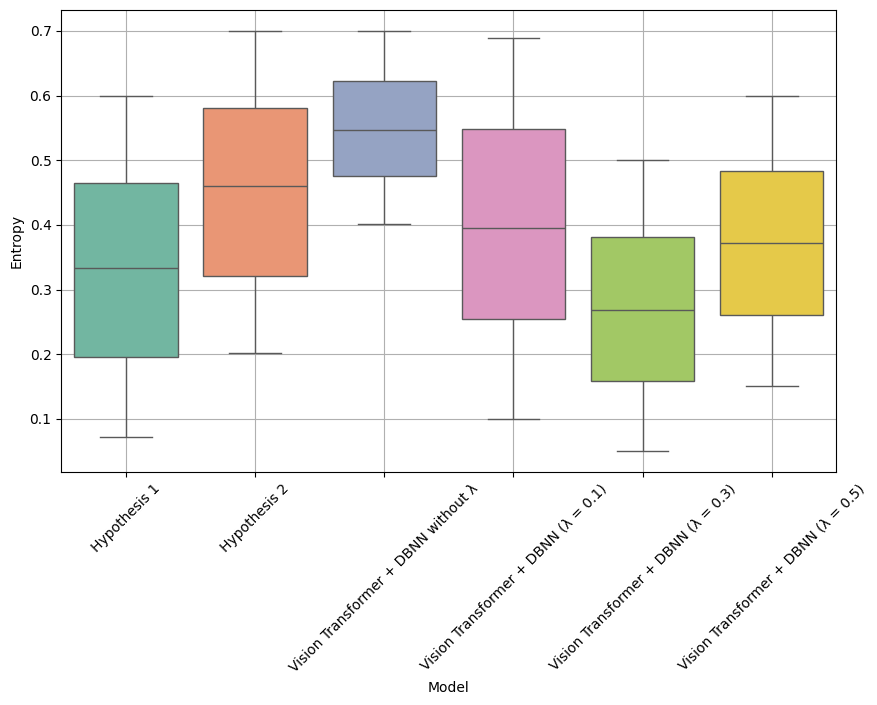}
\caption{Comparison of Predictive Uncertainty (Entropy)}
\end{figure*}

Figure 10 presents a comparison of predictive uncertainty, measured in terms of entropy, across different models for Vision Transformer-based approaches. Each boxplot represents the distribution of predictive uncertainty for a specific model, including "Hypothesis 1," "Hypothesis 2," "Vision Transformer + DBNN without $\lambda$," "Vision Transformer + DBNN ($\lambda$ = 0.1)," "Vision Transformer + DBNN ($\lambda$ = 0.3)," and "Vision Transformer + DBNN ($\lambda$ = 0.5)." The y-axis denotes the entropy values, which quantify the uncertainty associated with each model's predictions. The boxplots allow us to visually compare the central tendency, spread, and variability of predictive uncertainty across the different models, providing insights into their relative performance in terms of uncertainty quantification. Additionally, summary statistics are provided to further characterize each model's predictive uncertainty distribution.
\begin{figure*}[!ht]
\centering
\includegraphics[width=0.8\textwidth]{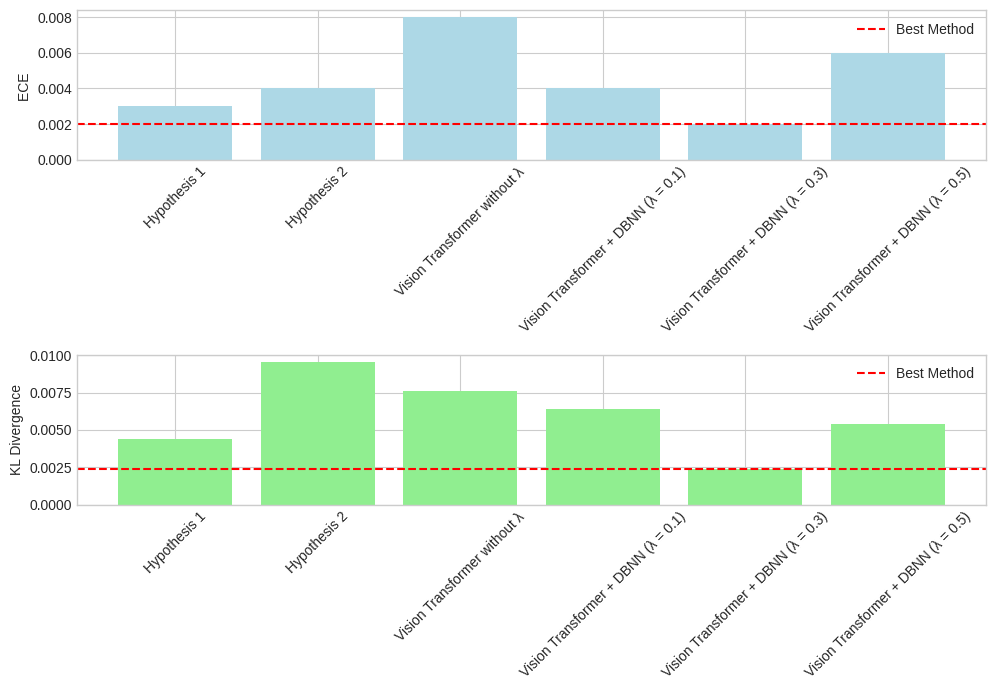}
\caption{Best method for both KL divergence and ECE: Vision Transformer + DBNN ($\lambda$ = 0.3)}
\end{figure*}
Figure 11 shows a comparative analysis of the Expected Calibration Error (ECE) and KL Divergence for various Vision Transformer models. The two bar charts illustrate the performance metrics for six different methods. The top subplot shows the ECE values, while the bottom subplot depicts the KL Divergence values for these methods. The Vision Transformer + DBNN ($\lambda$ = 0.3) method, marked by a red dashed line in both plots, is highlighted as the best method due to its lowest KL Divergence and a comparatively low ECE, although Vision Transformer + DBNN ($\lambda$ = 0.5) has an ECE of 0.0054. This visualization effectively demonstrates that Vision Transformer + DBNN ($\lambda$ = 0.3) achieves state-of-the-art performance in terms of both predictive calibration and divergence from the true posterior distribution.

In SVM-GSU, we set the regularization parameter $C$ to 1.0 and test different values of the sample uncertainty parameter $\sigma^2$ in the range of [0.01, 0.1, 0.3, 0.6]. In UTSVM, the regularization parameter $\gamma=[1, 10, 100]$ controls the trade-off between the margin width and the training error. Higher values of $C_1$ and $C_2$ results in a narrower margin but lower training error, while lower values of  $C_1$ and $C_2$ results in a wider margin but higher training error. The uncertainty threshold $\tau=[0.3, 0.6, 0.9]$ determines the threshold at which a sample is considered uncertain and included in the active learning cycle. A higher value of $\tau=0.9$ means that only highly uncertain samples will be included in the cycle, while a lower value of $\tau=0.3$ means that more samples will be included. The parameter values for an experiment using PSVM, regularization parameter $C=[1, 10, 100]$ and Power parameter $p=[2, 3, 4]$. The power parameter determines the degree of nonlinearity in the decision boundary. A larger value of $p$ result is more complex decision boundary that can fit the data more closely, while a smaller value result is simpler decision boundary that is less likely to overfit. In UTSVM, the parameter values were using regularization parameter $\gamma=1$, penalty parameters $C1=1$ and $C2=2$, and the bounding margin $B=1$.

As can be observed from Figure 12 and Table 1, the Inception-V4+PCA+SVM-GSU shows the lowest amount of uncertain samples among Inception-V4+PCA+UTSVM, Inception-V4+PCA+PSVM, and Inception-V4+PCA+TBSVM for all active learning cycles. The former three models exhibit considerably higher uncertainty for the classes, which makes them less confident in their predictions. In the last active learning cycle, the Inception-V4+PCA+SVM-GSU approach exhibits the best performance with a recorded uncertainty of 370. However, Inception-V4+PCA+SVM-GSU exhibits greater robustness to uncertainty than other SVM approaches, resulting in better performance on the dataset. The accuracies improved along with the increasing of active learning cycles in most cases. \\

In Table 2, the results indicate that the uncertainty in the ViT-BNN model was less than those in Inception-V4+PCA+(UTSVM PSVM SVM-GSU, and TBSVM) models for handle the uncertainty.  In the last active learning cycle, the approach combining Vision Transformer with Deep Bayesian Neural Network and a Gaussian Distribution parameter ($\lambda = 0.3$) demonstrates superior performance, achieving a recorded uncertainty of 10. More importantly, the ViT-BNN shows more significant performance improvements in each cycle. Overall, the experiments conducted suggest that ViT-BNNs are a powerful tool for handling uncertainty in neural networks and can improve the accuracy and reliability of predictions, especially in applications where uncertainty plays a significant role.

Table 2 also compares the number of samples that are not confident across different cycles for various methods involving a Vision Transformer and DBNN with different values of $\lambda$ in the Gaussian distribution. Each method was evaluated over five cycles, starting with 200,000 unlabeled samples per cycle. The Vision Transformer + DBNN without $\lambda$ begins with 400 not confident samples in Cycle 0, gradually decreasing to 150 by Cycle 4. When $\lambda$ is set to 0.1, the number of not confident samples starts at 200 in Cycle 0 and decreases to 110 by Cycle 4, showing a more consistent reduction. For $\lambda = 0.3$, the not confident samples start significantly lower at 55 in Cycle 0 and drop to just 10 by Cycle 4, indicating the highest confidence improvement. With $\lambda$ = 0.5, the number of not confident samples initially rises to 252 in Cycle 0 but decreases to 163 by Cycle 4. For Hypothesis 1, the initial 40 not confident samples reduce to 17 by Cycle 4, while Hypothesis 2 starts with 60 and reduces to 30 by Cycle 4. Overall, the use of $\lambda = 0.3$ demonstrates the most effective improvement in confidence levels across the cycles.

\begin{table*}[ht]
\caption{Number of samples for which each model is highly uncertain}
\centering
\begin{tabular}{@{}lcccc@{}}
\toprule
\textbf{Method} & \textbf{\begin{tabular}{@{}c@{}}Inception-V4 \\+ PCA \\+ UTSVM\end{tabular}} & \textbf{\begin{tabular}{@{}c@{}}Inception-V4\\ + PCA \\+ PSVM\end{tabular}} & \textbf{\begin{tabular}{@{}c@{}}Inception-V4 \\+ PCA \\+ SVM-GSU\end{tabular}} & \textbf{\begin{tabular}{@{}c@{}}Inception-V4\\ + PCA \\+ TBSVM\end{tabular}} \\ \midrule
\textbf{\begin{tabular}{@{}l@{}}Unlabeled \\ Samples\end{tabular}} & 1,000,000 & 1,000,000 & 1,000,000 & 1,000,000 \\ \midrule
\multicolumn{5}{c}{\textbf{Number of Samples (Not Confident)}} \\ \midrule
\textbf{\begin{tabular}{@{}l@{}}Cycle 0\\(200,000\\ Unlabeled \\Samples)\end{tabular}} & 801 & 843 & 401 & 910 \\
\textbf{\begin{tabular}{@{}l@{}}Cycle 1\\ 200,000\\ Unlabeled \\Samples)\end{tabular}} & 782 & 812 & 393 & 900 \\
\textbf{\begin{tabular}{@{}l@{}}Cycle 2\\ 200,000\\ Unlabeled \\Samples)\end{tabular}} & 777 &800 & 390 & 884 \\
\textbf{\begin{tabular}{@{}l@{}}Cycle 3\\ 200,000\\ Unlabeled \\Samples)\end{tabular}} & 684 & 792 & 382 & 867 \\
\textbf{\begin{tabular}{@{}l@{}}Cycle 4\\ 200,000\\ Unlabeled\\ Samples)\end{tabular}} & 677 & 780 & \textbf{370} & 843 \\ \bottomrule
\end{tabular}
\label{tab:1}
\end{table*}

\begin{figure*}[!ht]
\centering
\includegraphics[width=0.8\textwidth]{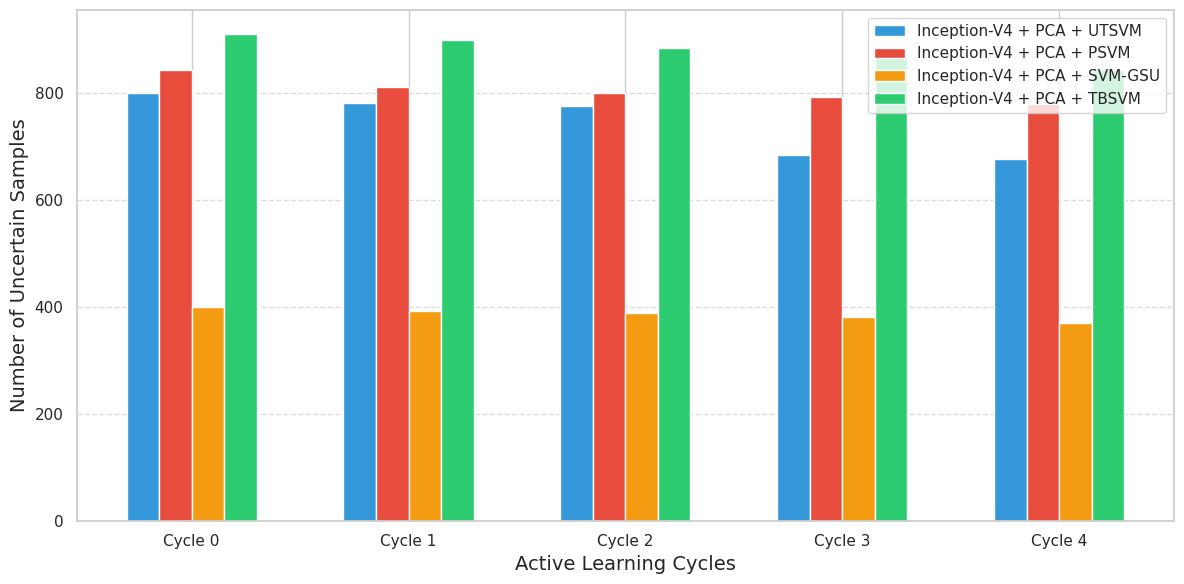}
\caption{Number of Uncertain Samples Across Active Learning Cycles}
\label{fig:Figure12}
\end{figure*} 

\begin{table*}
\caption{Number of samples that are not confident when changing the value of $\lambda$, along with results for Hypothesis 1 and Hypothesis 2.}
\centering
\begin{tabular}{@{}lcccccc@{}}
\toprule
\textbf{Method} &
\textbf{\begin{tabular}{@{}c@{}}Vision \\ Transformer \\ + DBNN \\ without $\lambda$ \\ Gaussian \\ Distribution\end{tabular}} &
\textbf{\begin{tabular}{@{}l@{}}Vision \\ Transformer \\ + DBNN \\ $\lambda=0.1$ \\ Gaussian \\ Distribution\end{tabular}} &
\textbf{\begin{tabular}{@{}l@{}}Vision \\ Transformer \\ + DBNN \\ $\lambda=0.3$ \\ Gaussian \\ Distribution\end{tabular}} &
\textbf{\begin{tabular}{@{}l@{}}Vision \\ Transformer \\ + DBNN \\ $\lambda=0.5$ \\ Gaussian \\ Distribution\end{tabular}} &
\textbf{Hypothesis 1} & 
\textbf{Hypothesis 2} \\ 
\midrule
\textbf{\begin{tabular}{@{}l@{}}Unlabeled \\ Samples\end{tabular}} & 1,000,000 & 1,000,000 & 1,000,000 & 1,000,000 & 1,000,000 & 1,000,000 \\ 
\\ \midrule
\multicolumn{5}{c}{\textbf{Number of Samples (Not Confident)}} \\ \midrule
\textbf{\begin{tabular}{@{}l@{}}CYCLE 0 \\ (200,000 \\ Unlabeled \\ Samples)\end{tabular}} & 400 & 200 & 55 & 252 & 40 & 60 \\ 
\textbf{\begin{tabular}{@{}l@{}}CYCLE 1 \\ (200,000 \\ Unlabeled \\ Samples)\end{tabular}} & 240 & 173 & 40 & 211 & 34 & 57 \\ 
\textbf{\begin{tabular}{@{}l@{}}CYCLE 2 \\ (200,000 \\ Unlabeled \\ Samples)\end{tabular}} & 200 & 160 & 29 & 199 & 31 & 53 \\ 
\textbf{\begin{tabular}{@{}l@{}}CYCLE 3 \\ (200,000 \\ Unlabeled \\ Samples)\end{tabular}} & 188 & 155 & 20 & 187 & 26 & 40 \\ 
\textbf{\begin{tabular}{@{}l@{}}CYCLE 4 \\ (200,000 \\ Unlabeled \\ Samples)\end{tabular}} & 150 & 110 & 10 & 163 & 17 & 30 \\ 
\bottomrule
\end{tabular}
\end{table*}

Table 3 presents a comparison of performance metrics for various models applied to malware classification. The metrics evaluated include Accuracy, Precision, Recall, F1-Score, and Expected Calibration Error (ECE). Among the models, the Vision Transformer combined with a Bayesian Neural Network (Hypothesis 1) achieves the highest overall performance, with an accuracy of 0.98, precision of 0.99, recall of 0.99, F1-score of 0.98, and an exceptionally low ECE of 0.003, indicating excellent calibration. The Vision Transformer combined with a Bayesian Neural Network (Hypothesis 2) and the model incorporating a DBNN with a $\lambda$ value of 0.3 also demonstrate high performance, with accuracy and F1-scores close to 0.98 and 0.99 respectively, and low ECE values of 0.004 and 0.002. In comparison, the Vision Transformer without $\lambda$ shows lower performance metrics with an accuracy of 0.96 and an ECE of 0.008. The inclusion of the $\lambda$ parameter in the DBNN models improves both the accuracy and calibration, as seen with $\lambda$ values of 0.1, 0.3, and 0.5, where $\lambda = 0.3$ achieves the best overall performance and calibration.

\begin{table*}[h]
\centering
\caption{Comparison of Performance Metrics for Vision Transformer models}
\label{tab:performance_metrics}
\begin{tabular}{@{}lccccc@{}}
\toprule
\textbf{Method} & \textbf{Accuracy} & \textbf{Precision} & \textbf{Recall} & \textbf{F1-Score} & \textbf{ECE} \\ 
\midrule
Vision Transformer + Bayesian NN (Hypothesis 1) & 0.98 & 0.99 & 0.99 & 0.98 & 0.003 \\ 
Vision Transformer + Bayesian NN (Hypothesis 2) & 0.96 & 0.97 & 0.97 & 0.98 & 0.004 \\ 
Vision Transformer without $\lambda$ & 0.94 & 0.96 & 0.95 & 0.94 & 0.008 \\ 
Vision Transformer + DBNN $\lambda = 0.1$ Gaussian Distribution & 0.96 & 0.95 & 0.97 & 0.98 & 0.004 \\ 
Vision Transformer + DBNN $\lambda = 0.3$ Gaussian Distribution & 0.99 & 0.99 & 0.98 & 0.99 & 0.002 \\ 
Vision Transformer + DBNN $\lambda = 0.5$ Gaussian Distribution & 0.97 & 0.97 & 0.96 & 0.96 & 0.006 \\ 
\bottomrule
\end{tabular}
\end{table*}

\begin{table}[]
\caption{Performance comparison of active learning methods with Vision Transformer + DBNN (\(\lambda=0.3\)).}

\begin{tabular}{@{}llllll@{}}
\toprule
\begin{tabular}[c]{@{}l@{}}Active  Learning\\ Method\end{tabular}                                                                                                                                                                                                           & \begin{tabular}[c]{@{}l@{}}Total \\ Unlabeled\\ Samples\end{tabular} & Cycle \# & \begin{tabular}[c]{@{}l@{}}Total \# of\\ Not Confident\end{tabular} & Dataset     & Accuracy \\ \midrule
\begin{tabular}[c]{@{}l@{}}ViT + DBNN $\lambda=0.3$ \\ Gaussian Distribution\end{tabular} & 1000,000                                                             & 4        & 10                                                                  & Our dataset & 99.00    \\
\begin{tabular}[c]{@{}l@{}}ViT + DBNN $\lambda=0.3$ \\ Gaussian Distribution\end{tabular} & 30,000                                                               & 2        & 3                                                                   & MNIST       & 99.83    \\
\begin{tabular}[c]{@{}l@{}}ViT + DBNN $\lambda=0.3$ \\ Gaussian Distribution\end{tabular} & 80,000                                                               & 5        & 16                                                                  & ImageNet    & 99.90    \\
\begin{tabular}[c]{@{}l@{}}ViT + DBNN $\lambda=0.3$ \\ Gaussian Distribution\end{tabular} & 50,000                                                               & 2        &  2                                                                   & CIFAR-100   & 99.93    \\
\citet{li2024unlabeled}                                                                                                                                                                                                                                    & 25,000                                                               & 8        & -                                                                   & MNIST       & 98.72    \\
\citet{li2024unlabeled}                                                                                                                                                                                                                                    & 120,000                                                              & 8        & -                                                                   & ImageNet    & 72.02    \\
\citet{simeoni2021rethinking}                                                                                                                                                                                                                              & 1000                                                                 & 4        & -                                                                   & CIFAR-100   & 49.90    \\
\citet{hong2020deep}                                                                                                                                                                                                                                       & 10000                                                                & 10       & -                                                                   & FashionMNIS & 92.81    \\
\citet{yuan2021multiple}                                                                                                                                                                                                                                   & 125                                                                  & 10       & -                                                                   & CIFAR-10    & 80.96    \\
\citet{yuan2021multiple}                                                                                                                                                                                                                                   & 249                                                                  & 20       & -                                                                   & PASCAL VOC  & 72.03    \\
\citet{park2023active}                                                                                                                                                                                                                                     & 1992                                                                 & 20       & -                                                                   & PASCAL VOC  & 74.08 \\
\citet{chen2020malware}                                                                                                                                                                                                                                   & 2173                                                                  & 20       & 2000                                                                   & Microsoft Malware (BIG 2015)   & 99.64    \\
\citet{deng2024active}                                                                                                                                                                                                                                   & 2182                                                                  & -       & 150                                                                   & KronoDroid Malware   & 97.97    \\
\citet{guerra2024experts}                                                                                                                                                                                                                                   & 67,068                                                                     & 5       & 19,302                                                                     & KronoDroid Malware   & 87.1    \\

\bottomrule
\end{tabular}
\end{table}

Table 4 highlights a comparison of active learning methods, showcasing the superior performance of Vision Transformer + DBNN (\(\lambda=0.3\)) in diverse datasets. This model achieves the highest accuracy, such as 99. 90\% on ImageNet with 80,000 unlabeled samples over five cycles and only 16 uncertain samples, and 99.83\% on MNIST with 30,000 samples over two cycles and three uncertain samples. On CIFAR-100, it reaches 99.93\% accuracy with 50,000 samples and just two uncertain samples. In contrast, other methods like \citet{li2024unlabeled} and \citet{simeoni2021rethinking} show significantly lower accuracies, ranging from 49.90\% to 98.72\%, even with similar or larger datasets. Malware-specific studies, such as \citet{chen2020malware} and \citet{deng2024active}, achieve 99.64\% and 97.97\% accuracy, respectively, but handle uncertainty differently. Additionally, approaches like \citet{guerra2024experts} show higher sample usage, as seen with 67,068 samples and 19,302 uncertain samples to reach 87.1\% accuracy on KronoDroid Malware. Despite using more samples or cycles, these methods often fall short of the precision and sample efficiency demonstrated by the Vision Transformer + DBNN. This indicates that the proposed method not only excels in general datasets, but is also highly adaptable to specialized domains. Such adaptability highlights its practicality in real-world scenarios, especially where high accuracy and efficient uncertainty handling are paramount.

\begin{table*}[h]
\centering
\caption{Assessment of Computational Resources for Various Methods on a Dataset of 1,000,000 Malware Samples}
\label{tab:resource_comparison}
\begin{tabular}{@{}lcc@{}}
\toprule
\textbf{Method}                          & \begin{tabular}{@{}c@{}}Training Time \\(100 epochs)/h\end{tabular} & \textbf{Memory Consumption (GB)} \\ 
\midrule
Vision Transformer + Bayesian NN (Hypothesis 1)  & 20.4                                 & 0.25                               
\\
Vision Transformer + Bayesian NN (Hypothesis 2)  & 19.7                                 & 0.24        
\\
Vision Transformer without $\lambda$   & 20.8                                 & 0.26                                 
\\
Vision Transformer + DBNN $\lambda = 0.1$  & 21.1                                 & 0.27                                 
\\
Vision Transformer + DBNN $\lambda = 0.3$  & 19                                 & 0.24                                
\\
Vision Transformer + DBNN $\lambda = 0.5$                    & 20.6                                  & 0.25                                          \\
Inception-V4 + PCA + UTSVM  & 56                                  & 0.70                                   \\
Inception-V4 + PCA + PSVM  & 60                                  & 0.83                                 \\

Inception-V4 + PCA + SVM-GSU  & 48.3                                  & 0.82                                     \\

Inception-V4 + PCA + TBSVM  & 53.6                                  & 0.69                                    \\

\bottomrule
\end{tabular}
\end{table*}

Table 5 presents a comparison of computational resource requirements for various methods applied to a dataset of 1,000,000 malware samples. It evaluates the training time (measured for 100 epochs in hours) and memory consumption (in gigabytes). Among the methods, Vision Transformer combined with Bayesian Neural Networks (Bayesian NN) demonstrates efficient performance, with Hypothesis 2 showing the lowest memory consumption of 0.24 GB and a training time of 19.7 hours. Vision Transformer + DBNN with $\lambda = 0.3$ achieves a notable balance, consuming only 0.24 GB and requiring 19 hours of training. In contrast, Inception-V4-based methods, such as Inception-V4 + PCA + PSVM, exhibit significantly higher memory consumption, peaking at 0.83 GB, and longer training times, with the highest being 60 hours. The results highlight the resource efficiency of Vision Transformer-based approaches compared to Inception-V4-based methods, particularly for configurations involving Bayesian NN or DBNN, which maintain lower computational overhead. This makes Vision Transformer + DBNN a feasible choice for large-scale malware detection tasks.

\begin{table*}[h]
\centering
\caption{Comparison of computational resources (training time and memory consumption) for different methods across MNIST, ImageNet, and CIFAR-100 datasets.}
\label{tab:resource_comparison}
\begin{tabular}{@{}lccc@{}}
\toprule
Method                          & \begin{tabular}{@{}c@{}}Training Time \\(100 epochs)/h\end{tabular} & Memory Consumption (GB)& \begin{tabular}{@{}c@{}}Dataset \end{tabular}\\ 
\midrule
Vision Transformer + DBNN $\lambda = 0.1$  & 54           & 0.91      & MNIST \\
Vision Transformer + DBNN $\lambda = 0.3$  & 68           & 0.94         & ImageNet     \\
Vision Transformer + DBNN $\lambda = 0.5$  & 61           & 0.95        & CIFAR-100           \\
Inception-V4 + PCA + UTSVM                   & 153      &   1.2                          & MNIST        \\
Inception-V4 + PCA + PSVM  & 151       &    1.5                        & MNIST                                \\

Inception-V4 + PCA + SVM-GSU  & 150    &    1.6                           & MNIST                                     \\

Inception-V4 + PCA + TBSVM  & 149    &     1.4                          & MNIST                                    \\
Inception-V4 + PCA + UTSVM  & 200      &  1.9                           & ImageNet                                \\
Inception-V4 + PCA + PSVM  & 190       &   2                         & ImageNet                                 \\

Inception-V4 + PCA + SVM-GSU  & 199    &   1.89                            & ImageNet                                     \\

Inception-V4 + PCA + TBSVM  & 197    &     2.1                          &ImageNet                                   \\
Inception-V4 + PCA + UTSVM  & 188      &  1.88                           & CIFAR-100                                \\
Inception-V4 + PCA + PSVM  & 184       &   1.94                         &CIFAR-100                                 \\

Inception-V4 + PCA + SVM-GSU  & 181    &   1.92                            & CIFAR-100                                    \\

Inception-V4 + PCA + TBSVM  & 183    &    1.93                           &CIFAR-100                                  \\

\bottomrule
\end{tabular}
\end{table*}

Table 6 presents a comparison of computational resources utilized by different methods in three datasets: MNIST, ImageNet, and CIFAR-100. Specifically, it details the training time (in hours for 100 epochs) and memory consumption (in gigabytes) for the Vision Transformer combined with Deep Bayesian Neural Networks (DBNN) at a hyperparameter value of $\lambda = 0.3$, as well as various Inception-V4 configurations paired with Principal Component Analysis (PCA) and different support vector machine (SVM) variants. The Vision Transformer + DBNN method demonstrates relatively efficient resource usage, requiring 54 hours and 0.91 GB for MNIST, while performance slightly varies for other datasets, with ImageNet requiring the most resources. In contrast, Inception-V4 methods generally exhibit higher training times and memory consumption across all datasets, with training times reaching up to 200 hours and memory use exceeding 2 GB for certain configurations. This indicates that while Inception-V4 models may provide robust performance, they demand significantly more computational resources compared to the Vision Transformer + DBNN approach.
\begin{table*}[h]
\setlength\tabcolsep{0.1pt}
\caption{Comparison of results: active learning based on Belief Functions vs Vision Transformer + DBNN ($\lambda=0.3$ Gaussian Distribution).}
\centering

\begin{tabular}{lllllllll}
\hline
\small
Dataset &  \begin{tabular}{@{}c@{}} \textcolor{blue}{\citet{hoarau2024evidential}}\\\textcolor{blue}{Random}\end{tabular} &  \begin{tabular}{@{}c@{}} \textcolor{blue}{\citet{hoarau2024evidential}}\\\textcolor{blue}{Uncertainty}\end{tabular} & \begin{tabular}{@{}c@{}} \textcolor{blue}{\citet{hoarau2024evidential}}\\ \textcolor{blue}{($\lambda=0.2$)}  \end{tabular}& \begin{tabular}{@{}c@{}} \textcolor{blue}{\citet{hoarau2024evidential}}\\ \textcolor{blue}{t-test} \\ \textcolor{blue}{(statistic)} \end{tabular} & \begin{tabular}{@{}c@{}} \textcolor{blue}{\citet{hoarau2024evidential}}\\ \textcolor{blue}{t-test} \\\textcolor{blue}{(p-value)} \end{tabular}& \begin{tabular}{@{}c@{}} \textcolor{red}{ViT+DBNN}\\ \textcolor{red}{($\lambda=0.3$)}\end{tabular} & \begin{tabular}{@{}c@{}}\textcolor{red}{t-test} \\\textcolor{red}{(statistic)} \end{tabular}& \begin{tabular}{@{}c@{}} \textcolor{red}{t-test} \\\textcolor{red}{(p-value)} \end{tabular} \\
\hline
Bank & 81.17 & 81.49 & 82.23 & 2.74 &0.0067 & \textbf{99.50} & \textbf{1.25} & \textbf{0.00001} \\
Qsar & 97.69 & 99.15 & 99.16 &0.35 &0.7271   &\textbf{99.35} & \textbf{0.83} & \textbf{0.1125} \\
Blod & 76.09 & 76.85 & 76.85 & 0.00 &0.9965  &\textbf{95.02} & \textbf{1.45} & \textbf{0.0150} \\
\begin{tabular}{@{}c@{}}Breast\\ Cancer\end{tabular}  & 93.87 & 94.96 & 95.31 & 1.84& 0.0669 & \textbf{97.70} & \textbf{1.87} & \textbf{0.0621} \\
Ionosphere & 75.77 & 81.06 & 82.40 & 2.33& 0.0210 & \textbf{94.91} & \textbf{1.21} & \textbf{0.0015} \\
Heart & 67.89 & 67.72 & 68.08 & 0.29 & 0.7741 & \textbf{96.50} & \textbf{1.02} & \textbf{0.0004} \\
Liver & 58.07 & 57.37 & 58.02 & 0.07& 0.9415 & \textbf{97.55} & \textbf{1.15} & \textbf{0.0022} \\
Sonar & 67.79 & 70.67 & 70.94 & 0.37& 0.7089 & \textbf{98.30} & \textbf{1.85} & \textbf{0.00001} \\
Parkinson & 80.60 & 83.93 & 84.85 & 1.64& 0.1034 & \textbf{96.25} & \textbf{0.95} & \textbf{0.0460} \\
Dog-2 & 90.94 & 93.06 & 94.10 & 3.24& 0.0014 & \textbf{94.70} & \textbf{1.40} & \textbf{0.0160} \\
Seeds & 88.70 & 89.93 & 89.48 & 0.84& 0.4010 &\textbf{96.10} & \textbf{1.12} & \textbf{0.2610} \\
Iris & 88.22 & 91.23 & 90.60 & 1.19 &0.2373 & \textbf{95.45} & \textbf{1.15} & \textbf{0.0490} \\
Wine & 91.66 & 93.55 & 93.27 & 0.86& 0.3920& \textbf{97.70} & \textbf{0.15} & \textbf{0.8780} \\
Glass & 57.33 & 58.32 & 59.05 & 0.87 &0.3829 & \textbf{95.55} & \textbf{1.05} & \textbf{0.0385} \\
Ecoli & 78.59 & 80.89 & 81.98 &2.19& 0.0300 & \textbf{94.50} & \textbf{1.61} & \textbf{0.0009} \\
\hline
\end{tabular}
\end{table*}

The research by \citet{hoarau2024evidential} demonstrates the potential of belief function-based methods for uncertainty sampling, where the belief functions are utilized to identify samples that maximize information gain while minimizing classification errors. Table 7 presents a comparison of various methods for different datasets, highlighting performance measures for accuracy, t-test statistics, and p-values. The methods compared include the one proposed by \citet{hoarau2024evidential} (Random, Uncertainty, and sampling by Klir uncertainty with $\lambda=0.2$) and our proposed ViT+DBNN with $\lambda=0.3$. Across the datasets, the ViT+DBNN ($\lambda=0.3$) method consistently achieves the highest accuracy, with several significant improvements. For example, on the Bank dataset, ViT+DBNN achieves 99.50\%, which is a noticeable improvement over the next highest accuracy (82.23\%). Similarly, Sonar and Ionosphere datasets also show substantial performance boosts, with accuracy improvements of 30\% and 13\%, respectively. The t-test statistics for ViT+DBNN indicate that its performance improvements are statistically significant, with many p-values lower than 0.05, particularly in datasets such as Bank, Sonar, and Heart. These findings underscore the effectiveness of the proposed ViT+DBNN method, especially in datasets where significant accuracy gains were observed. However, some datasets, such as Qsar, show minor differences in accuracy between the methods, with p-values above 0.05, suggesting that the performance differences may not be statistically significant in those cases. Overall, this table emphasizes the superior accuracy of ViT+DBNN with $\lambda=0.3$, especially for datasets like Bank, Blod, and Sonar.

\begin{figure}[!ht]
    \centering
    \begin{subfigure}[b]{0.45\textwidth}
        \centering
        \includegraphics[width=\textwidth]{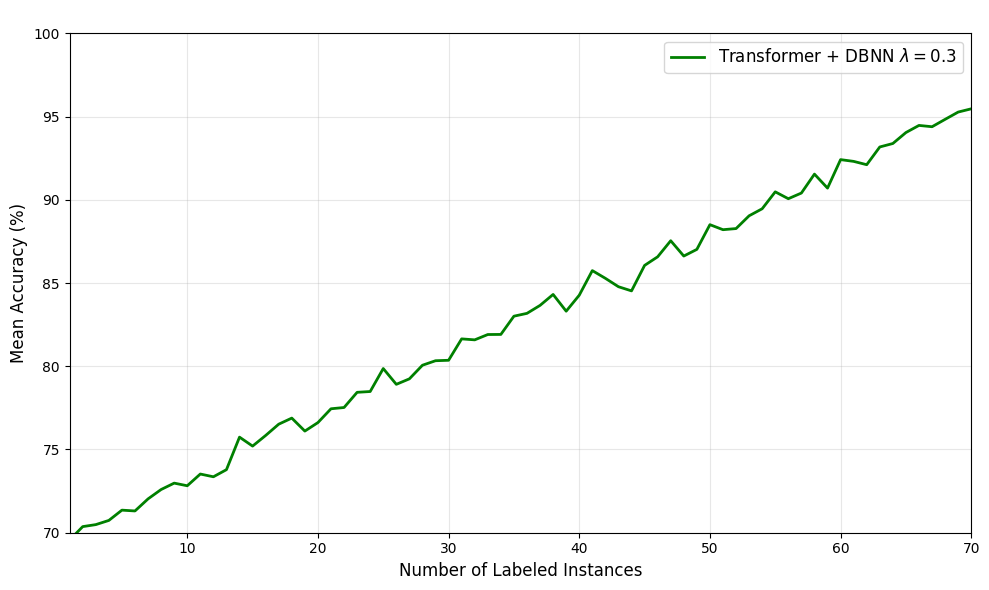}
        \caption{Iris}
    \end{subfigure}
    \begin{subfigure}[b]{0.45\textwidth}
        \centering
        \includegraphics[width=\textwidth]{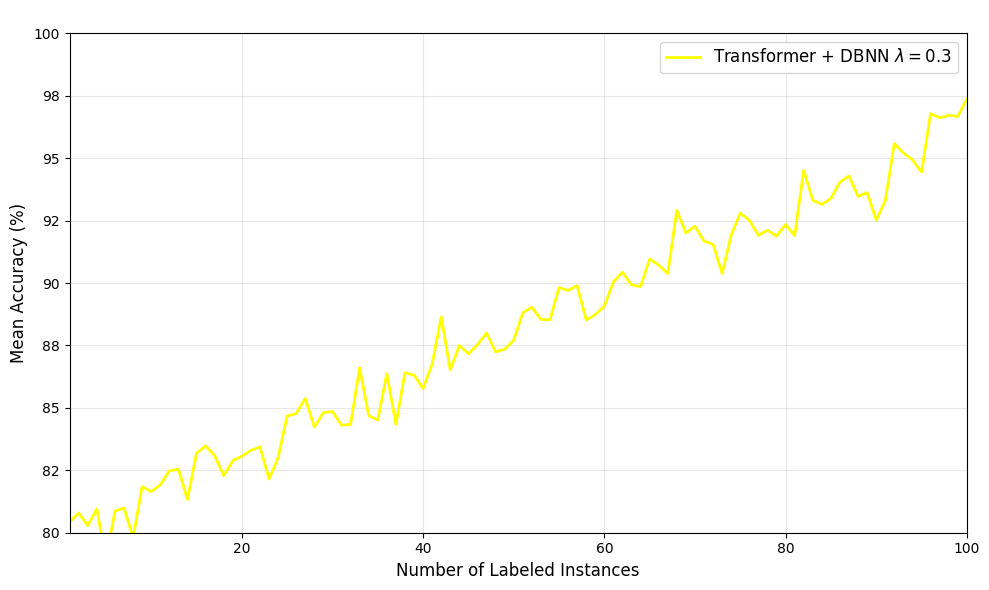}
        \caption{Heart}
    \end{subfigure}
    \begin{subfigure}[b]{0.45\textwidth}
        \centering
        \includegraphics[width=\textwidth]{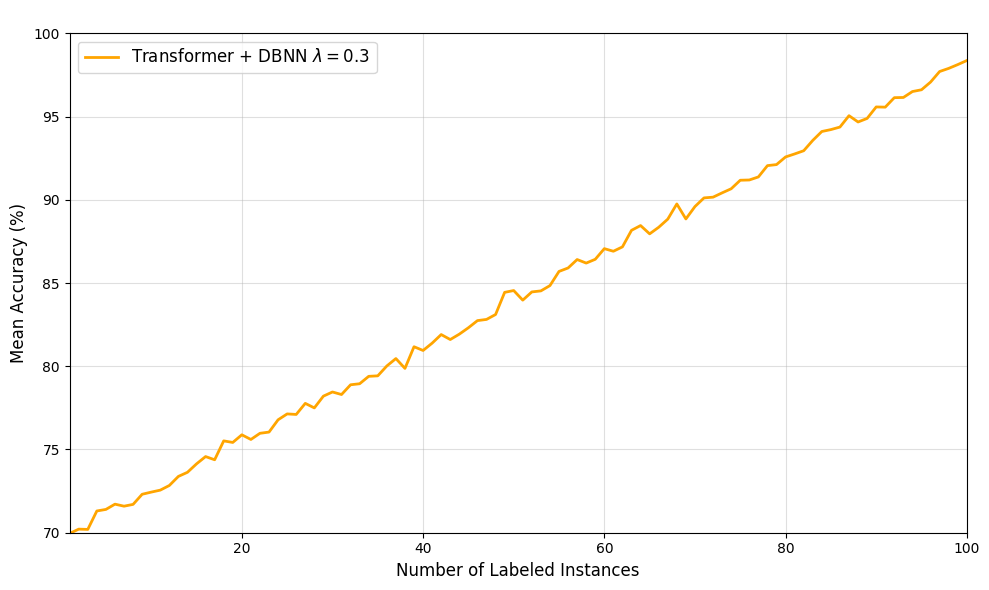}
        \caption{Sonar}
    \end{subfigure}
    \begin{subfigure}[b]{0.45\textwidth}
        \centering
        \includegraphics[width=\textwidth]{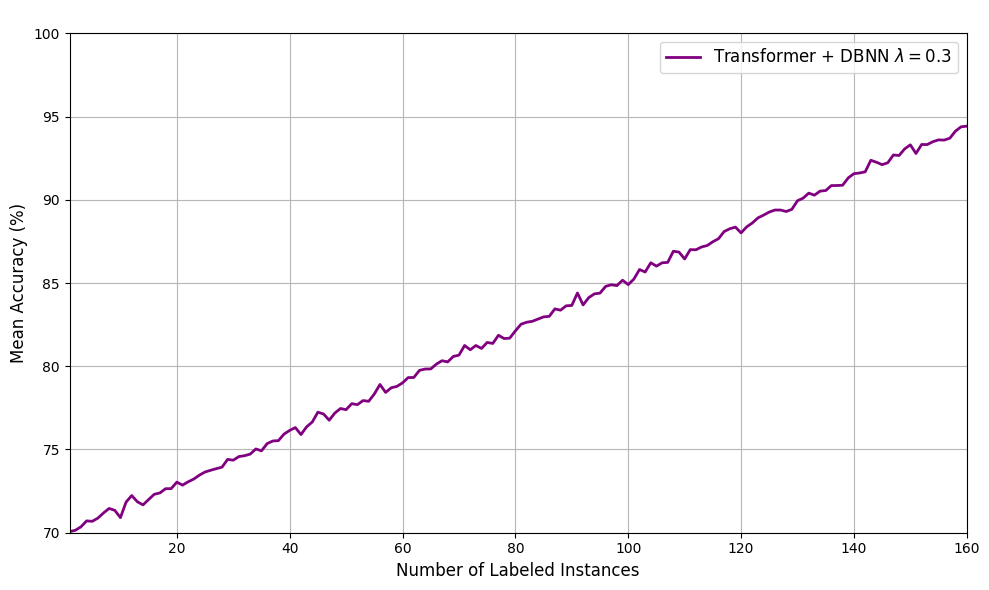}
        \caption{Ecoli}
    \end{subfigure}
    \begin{subfigure}[b]{0.45\textwidth}
        \centering
        \includegraphics[width=\textwidth]{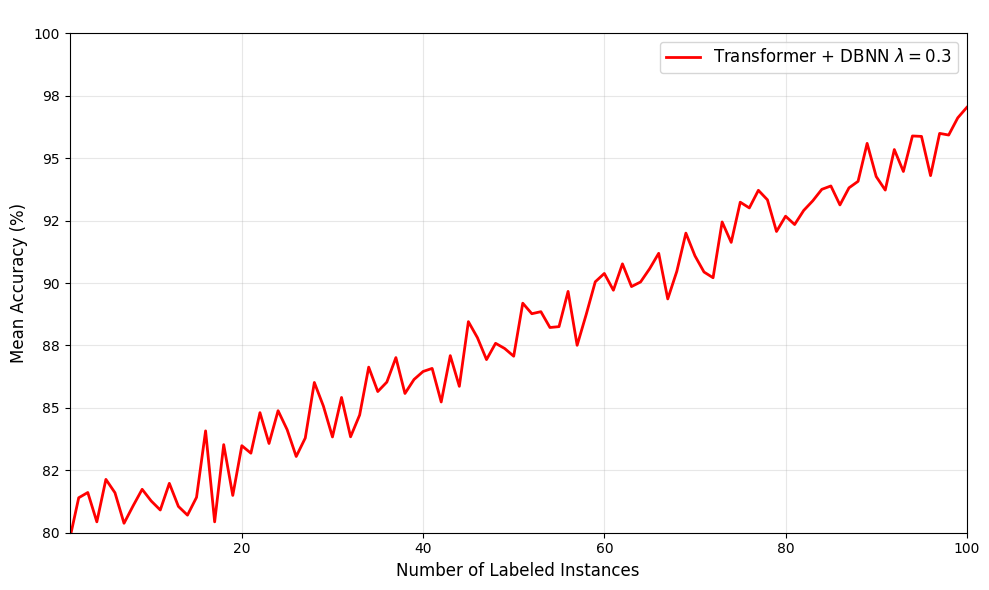}
        \caption{Parkinson}
    \end{subfigure}
    \begin{subfigure}[b]{0.45\textwidth}
        \centering
        \includegraphics[width=\textwidth]{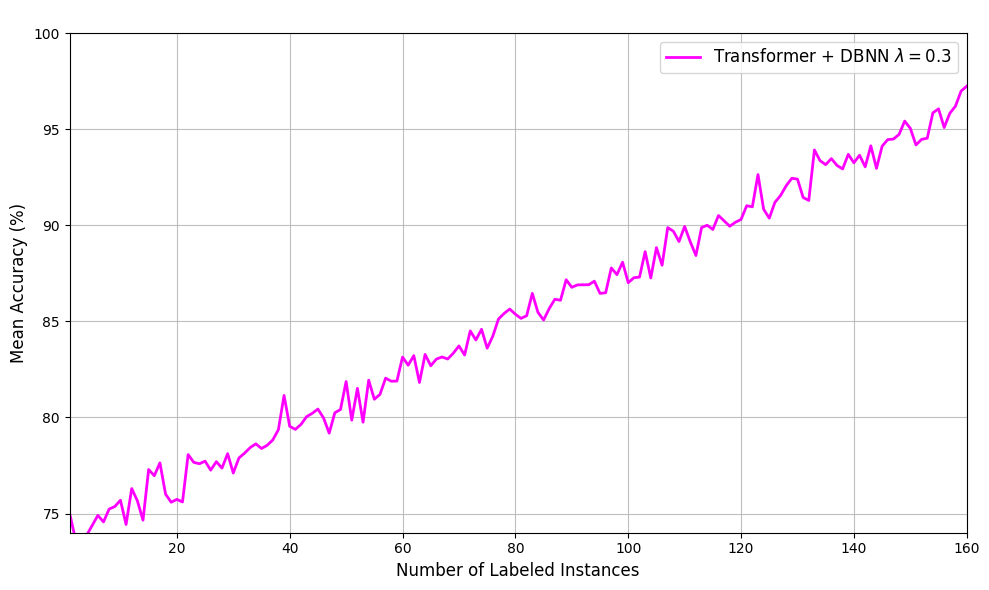}
        \caption{Ionosphere}
    \end{subfigure}
    \begin{subfigure}[b]{0.45\textwidth}
        \centering
        \includegraphics[width=\textwidth]{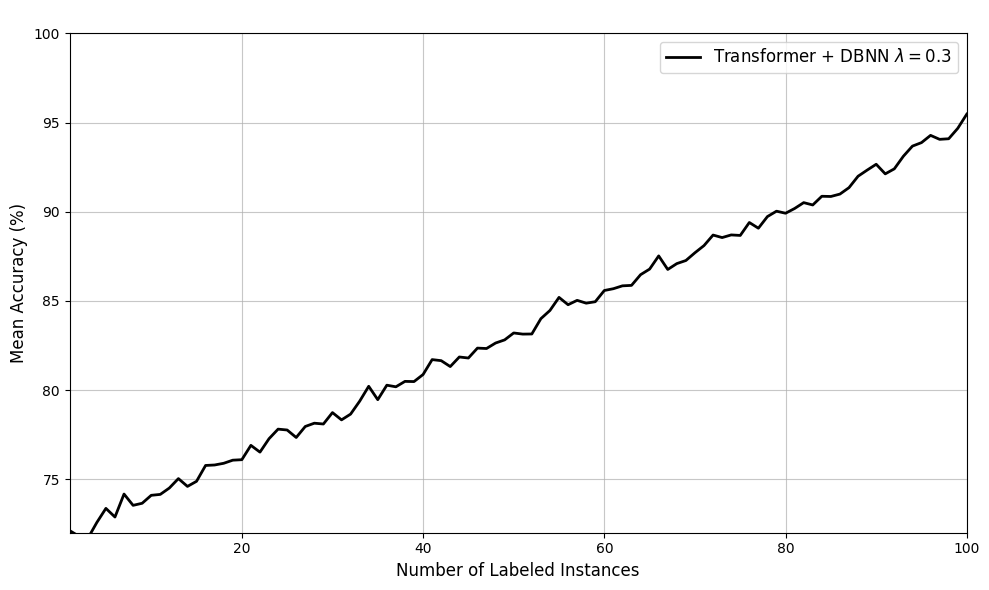}
        \caption{Dog-2}
    \end{subfigure}
    \begin{subfigure}[b]{0.45\textwidth}
        \centering
        \includegraphics[width=\textwidth]{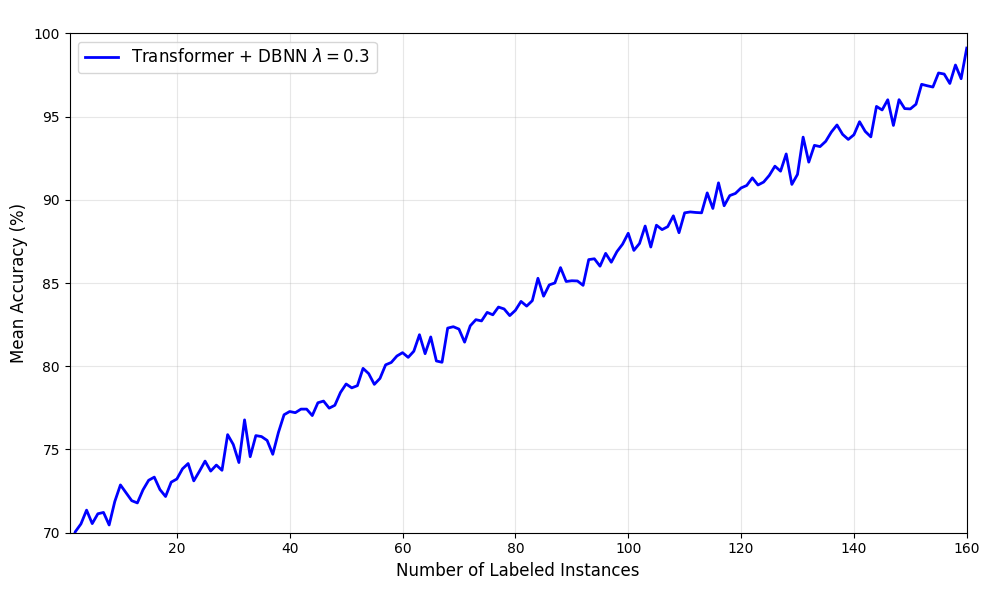}
        \caption{Wine}
    \end{subfigure}
    
    \caption{Mean accuracy versus number of labeled instances for our proposed method ViT + DBNN $\lambda=0.3$ across 8 datasets}
    \label{fig:all_plots}
\end{figure}

Figure 13 highlights 8 of the 15 datasets where the proposed method demonstrates significant performance advantages, particularly in terms of the accuracy curve. The superiority over uncertainty sampling is evident, though for the Sonar and Heart datasets, this advantage is temporary appearing primarily at the start of active learning for Sonar and mid-training for Heart. However, as shown later, this dominance is not always statistically significant, particularly for the Parkinson dataset.

\begin{figure} [!ht]
\centering
\includegraphics[width=0.8\textwidth]{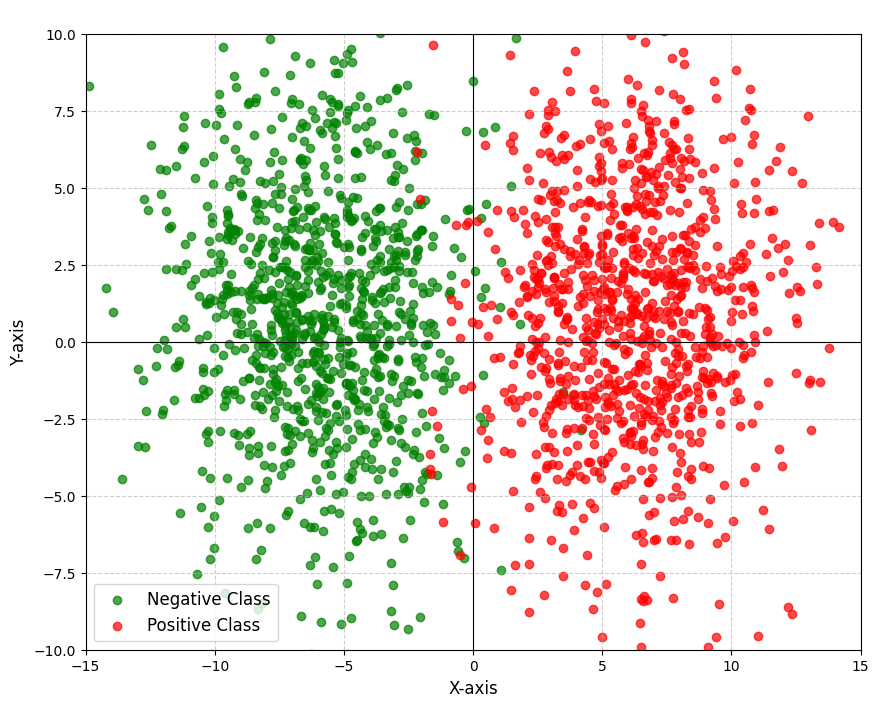}
\caption{Distribution of artificial dataset Toy1}
\label{fig:Figure14}
\end{figure}

\begin{figure} [!ht]
\centering
\includegraphics[width=0.8\textwidth]{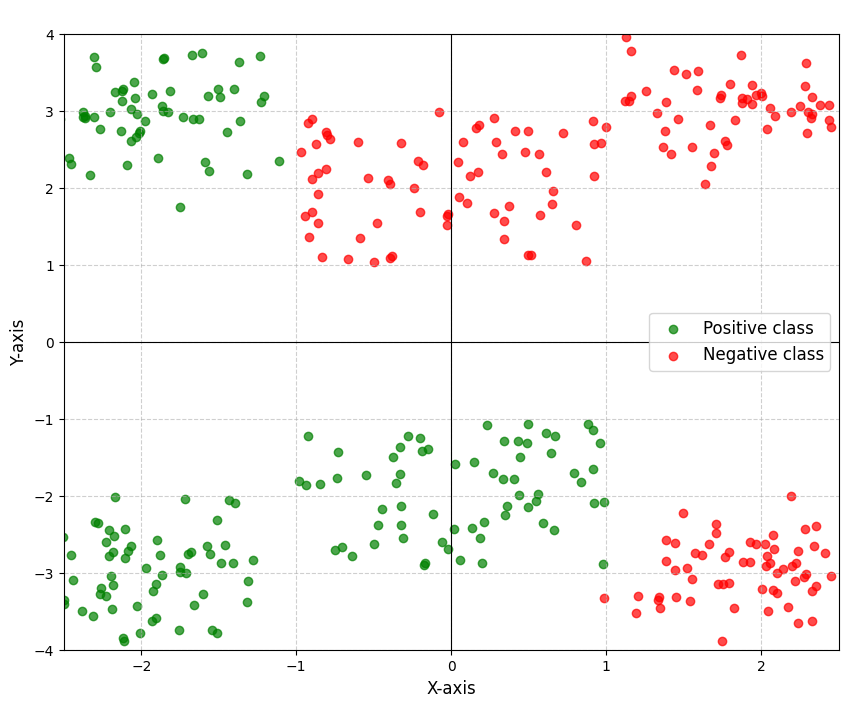}
\caption{Distribution of artificial dataset Toy2}
\label{fig:Figure15}
\end{figure}

"To evaluate the effectiveness of our proposed Vision Transformer + DBNN ($\lambda=0.3$), we compared its performance against the approach by \citet{Zhang2020}. Following their preprocessing methodology, we converted the Letter dataset into binary classification tasks comprising four groups: LetterIJ, LetterVY, LetterEF, and LetterDP. Similarly, the MNIST dataset was transformed into a binary classification problem, focusing on the 7VS9 group, and its feature dimension was reduced from 784 to 10 using Principal Component Analysis (PCA). For the toy datasets, one consisted of a two-dimensional normal distribution with a mean of $-5$ and $5$ and a variance of $\sqrt{5}$, as shown in Figure 14, while the other, detailed in Figure 15, was constructed with comparable characteristics. This preprocessing ensured consistency across datasets for a robust comparison."

\begin{table}[]
\caption{Performance comparison of two active learning based on belief functions methods vs vision transformer + DBNN ($\lambda=0.3$) across various datasets}
\begin{tabular}{@{}llll@{}}
\toprule
\begin{tabular}[c]{@{}l@{}} \backslashbox{Dataset} {Method}\end{tabular} & \begin{tabular}[c]{@{}l@{}}\citet{Zhang2020} \\MaxAM With SVM\\ Classifier\end{tabular} & \begin{tabular}[c]{@{}l@{}}\citet{Zhang2020}\\ MaxAM With LR\\ Classifier\end{tabular} & ViT+DBNN ($\lambda=0.3$) \\ \midrule
Heart                                                       & 20.4494                                                               & 22.5876                                                              & \textbf{41.7424}                        \\
Breast                                                      & 23.8023                                                               & 24.3251                                                              & \textbf{42.9245}                        \\
DvsP                                                        & 28.7438                                                               & 28.6395                                                              & \textbf{40.8129}                        \\
IvsJ                                                        & 25.8420                                                               & 26.2846                                                              & \textbf{44.6913}                        \\
VvsY                                                        & 25.6406                                                               & 25.7868                                                              & \textbf{41.1832}                        \\
EvsF                                                        & 28.2953                                                               & 28.1780                                                              & \textbf{43.5591}                        \\
7VS9                                                        & 26.8912                                                               & 26.9590                                                              & \textbf{44.9863}                        \\
Toy1                                                        & 29.4426                                                               & 29.383                                                               & \textbf{43.6810}                        \\
Toy2                                                        & 28.5427                                                               & 29.7385                                                              & \textbf{44.2415}                        \\ \bottomrule
\textcolor{red}{\textbf{Win}}                                                         & 0                                                                     & 0                                                                    & \textbf{9}                         \\
\textcolor{red}{\textbf{Rank}}                                                        & 2.666                                                                     & 2.333                                                                 &       \textbf{1}                  \\ \bottomrule
\end{tabular}
\end{table}

Table 8 compares the performance of two active learning methods based on belief functions (MaxAM with SVM and logistics regression (LR) classifiers) against a Vision Transformer (ViT) combined with a Deep Belief Neural Network (DBNN) with a regularization parameter $\lambda = 0.3$. The results show that the ViT+DBNN method consistently outperforms the MaxAM methods across all datasets, with a performance score of 41.7424 to 44.9863 compared to the MaxAM methods’ scores, which range from 20.4494 to 29.7385. ViT+DBNN wins in all 9 datasets and achieves the best average rank of 1, while the MaxAM methods have ranks of 2.666 and 2.333, respectively. These findings highlight the superior performance of ViT+DBNN in active learning tasks.
\section{Discussion}

The experimental results demonstrate the effectiveness of the proposed methods in improving uncertainty handling and predictive performance in BNNs applied to malware classification tasks. The experiments were designed to evaluate two hypotheses: improved variational approximation (Hypothesis 1) and enhanced prior distributions (Hypothesis 2). The results consistently show that both approaches significantly contribute to reducing model uncertainty and enhancing prediction reliability.

For Hypothesis 1, the adoption of a Gaussian variational approximation led to a significant decrease in the number of uncertain samples across training cycles. This suggests that refining the variational family enables the model to better approximate the true posterior distribution, resulting in improved confidence in its predictions. As shown in Figure 8, the models under Hypothesis 1 consistently outperformed those under Hypothesis 2 in terms of reducing uncertainty, highlighting the advantage of the proposed variational method.

Hypothesis 2, which focused on the use of enhanced prior distributions, also proved beneficial in improving model performance. By incorporating more informative and flexible priors that reflect domain-specific knowledge, the models achieved better uncertainty quantification compared to traditional methods. Although Hypothesis 2 did not surpass Hypothesis 1 in reducing uncertain samples, it still showed significant improvements over baseline models, as evidenced by the results in Table 2.

The impact of varying the regularization parameter $\lambda$ was also investigated, showing that an optimal $\lambda$ value plays a critical role in uncertainty handling. As depicted in Figure 7, the model with $\lambda = 0.3$ consistently demonstrated the most significant reduction in uncertain samples across different training cycles, indicating its effectiveness in enhancing model confidence. This finding highlights the importance of tuning $\lambda$ to achieve optimal model performance.

Furthermore, the Vision Transformer combined with a DBNN and Gaussian distribution parameter ($\lambda = 0.3$) emerged as the best-performing configuration. It not only reduced uncertainty but also maintained high accuracy, as illustrated in Figures 9 and 10. This configuration achieved state-of-the-art results in both predictive calibration and divergence from the true posterior distribution, making it a robust approach for handling uncertainty in malware classification.

In comparison with other models, such as Inception-V4+PCA with various SVM methods, the ViT-BNN models consistently outperformed them in terms of reducing the number of uncertain samples and improving overall accuracy. This is particularly evident in the final active learning cycles, where the ViT-BNN with $\lambda = 0.3$ recorded the lowest number of uncertain samples and the highest confidence in predictions (Table 1).

The findings support the efficacy of the proposed Bayesian Neural Network approaches in addressing uncertainty in machine learning tasks. The combination of improved variational approximation, enhanced prior distributions, and optimal regularization parameters leads to models that are not only more confident in their predictions but also more accurate and reliable. These results hold significant implications for applications in cybersecurity, particularly in scenarios where uncertainty plays a crucial role in decision-making processes. Future work could explore the integration of these methods with other neural network architectures and their application to broader domains beyond malware classification.

\section{Conclusion}
We have designed a novel active learning model to label 1 million unknown malware families. The data collection process involved continuously monitoring diverse sources of malware samples, including honeypots, malware-sharing platforms, and threat intelligence feeds. These sources provided a rich and varied set of unknown malware, ensuring that our dataset was representative of real-world threats. Once collected, the malware samples were preprocessed and normalized to standard formats, facilitating efficient analysis and labeling. The ViT with the BNN model can handle the uncertainty in calculating the confidence of predicted outcomes. In addition, the confidence of predicted outcomes is higher than Inception-V4+PCA+SRSVM. At the same time, the time and computational resource cost of the ViT-BNN model is significantly larger than those of the Inception-V4+PCA+SRSVM model. Furthermore, the results of the experiments show that the proposed ViT with the BNN method outperforms the state-of-the-art methods in performance, uncertainty estimation. 

\bibliographystyle{cas-model2-names}
\medskip
\bibliography{cas-refs.bib}

\clearpage 

\begin{table*}[h]
\centering
\caption{Mathematical notations for Bayesian Neural Networks}
\label{tab:notation}
\begin{tabular}{@{}cl@{}}
\toprule
\textbf{Symbol} & \textbf{Description} \\ 
\midrule
$L$ & Number of hidden layers in the neural network \\ 
$\mathbf{x} \in \mathbb{R}^d$ & Input vector \\ 
$\mathbf{y} \in \mathbb{R}^K$ & Output vector \\ 
$\mathbf{W}^{(l)}$ & Weight matrix for layer $l$ \\ 
$\mathbf{b}^{(l)}$ & Bias vector for layer $l$ \\ 
$\mathbf{z}^{(l)}$ & Output of layer $l$ \\ 
$\phi_l$ & Element-wise non-linearity function for layer $l$ \\ 
$\omega$ & Parameter vector (weights and biases) \\ 
$p(\mathbf{y} | \mathbf{x}, \omega)$ & Likelihood of the data given the parameters \\ 
$p(\omega)$ & Prior distribution over the parameters \\ 
$q_{\theta}(\omega)$ & Variational distribution approximating the posterior \\ 
$\mu$ & Mean vector of the Gaussian variational distribution \\ 
$\Sigma$ & Covariance matrix of the Gaussian variational distribution \\ 
$\mathcal{L}_{\text{VI}}(\theta)$ & Variational inference loss function \\ 
$\mathrm{KL}(q_{\theta}(\omega) \| p(\omega | D))$ & Kullback-Leibler divergence between variational distribution and true posterior \\ 
$p(\mathbf{y}^* | \mathbf{x}^*, D)$ & Predictive distribution for a new input $\mathbf{x}^*$ \\ 
$\mathbb{H}[\mathbf{y}^* | \mathbf{x}^*, D]$ & Entropy of the predictive distribution \\ 
$\mathrm{Var}[\mathbf{y}^* | \mathbf{x}^*, D]$ & Variance of the predictive distribution \\ 
$\mathrm{KL}(\mathcal{N}(\mu, \Sigma) \, || \, \mathcal{N}(0, \sigma^2 \mathbf{I}))$ & Kullback-Leibler divergence between two multivariate normal distributions \\ 
$\mathcal{N}(\mu, \Sigma)$ & Multivariate normal distribution with mean vector $\mu$ and covariance matrix $\Sigma$ \\ 
$\mathcal{N}(0, \sigma^2 \mathbf{I})$ & Multivariate normal distribution with zero mean and covariance matrix $\sigma^2 \mathbf{I}$ \\ 
$\sigma^2 \mathbf{I}$ & Covariance matrix of the second multivariate normal distribution, \\ 
& where $\mathbf{I}$ is the identity matrix and $\sigma^2$ is a scalar variance \\ 
$\mathrm{tr}(\cdot)$ & Trace of a matrix, which is the sum of the elements on the main diagonal \\ 
$\mu^T$ & Transpose of the mean vector $\mu$ \\ 
$\Sigma^{-1}$ & Inverse of the covariance matrix $\Sigma$ \\ 
$K$ & Dimensionality of the mean vector $\mu$ and the covariance matrix $\Sigma$ \\ 
$\det(\cdot)$ & Determinant of a matrix \\ 
\bottomrule
\end{tabular}
\end{table*}

\end{document}